\documentclass[12pt,preprint]{aastex}

\usepackage{color}

\newcommand{\sns}{SDSS-II Supernova Survey}

\newcommand{\avgn}{\langle N \rangle}
\newcommand{\nobs}{N_{\mathrm{SNe}}}
\newcommand{\rateunits}{\mathrm{SNe}~\mathrm{yr}^{-1}~\mathrm{Mpc}^{-3}~h_{70}^{3}}
\newcommand{\nsnetotall}{774}
\newcommand{\nsnetotallconf}{312}

\newcommand{\nsneconf}{270}
\newcommand{\nsneconfpct}{52}
\newcommand{\nsnephoto}{246}
\newcommand{\nsnephotoz}{113}
\newcommand{\nsnephotonoz}{133}

\newcommand{\nsnezltohthree}{516}

\newcommand{\nnoniaall}{87}
\newcommand{\nnoniaonetwo}{42}
\newcommand{\nnoniathree}{45}

\newcommand{\lowzval}{2.35}
\newcommand{\lowzvalhistat}{0.45}
\newcommand{\lowzvalhisyst}{0.15}
\newcommand{\lowzvallowstat}{0.39}
\newcommand{\lowzvallowsyst}{0.003}

\newcommand{\lowzvalonefive}{2.69}
\newcommand{\lowzvalonefivehistat}{0.34}
\newcommand{\lowzvalonefivehisyst}{0.21}
\newcommand{\lowzvalonefivelowstat}{0.30}
\newcommand{\lowzvalonefivelowsyst}{0.01}

\newcommand{\lowzvaloheight}{2.93}
\newcommand{\lowzvalhistatoheight}{0.90}
\newcommand{\lowzvalhisystoheight}{0.71}
\newcommand{\lowzvallowstatoheight}{0.17}
\newcommand{\lowzvallowsystoheight}{0.04}

\newcommand{\zoff}{0.21}

\newcommand{\apval}{3.43}
\newcommand{\apvalhi}{0.15}
\newcommand{\apvallo}{0.15}

\newcommand{\nuval}{2.04}
\newcommand{\nuvalhi}{0.90}
\newcommand{\nuvallo}{0.89}
\newcommand{\powcorr}{-0.019}

\newcommand{\apvalav}{4.38}
\newcommand{\apvalhiav}{0.20}
\newcommand{\apvalloav}{0.19}

\newcommand{\nuvalav}{4.66}
\newcommand{\nuvalhiav}{0.93}
\newcommand{\nuvalloav}{0.92}

\begin{document}

\title{Measurements of the Rate of Type Ia Supernovae at Redshift 
$\lesssim 0.3$ from the SDSS-II Supernova Survey
}

\date{\today}

\email{bdilday@physics.rutgers.edu}

\author{
Benjamin~Dilday,\altaffilmark{1,2,3}
Mathew~Smith,\altaffilmark{4,5}
Bruce~Bassett,\altaffilmark{4,6}
Andrew~Becker,\altaffilmark{7}
Ralf~Bender,\altaffilmark{8,9}
Francisco~Castander,\altaffilmark{10}
David~Cinabro,\altaffilmark{11}
Alexei~V.~Filippenko,\altaffilmark{12}
Joshua~A.~Frieman,\altaffilmark{13,14}
Llu\'{\i}s~Galbany,\altaffilmark{15}
Peter~M.~Garnavich,\altaffilmark{16}
Ariel~Goobar,\altaffilmark{17,18}
Ulrich~Hopp,\altaffilmark{8,9}
Yutaka~Ihara,\altaffilmark{19}
Saurabh~W.~Jha,\altaffilmark{1}
Richard~Kessler,\altaffilmark{3,13}
Hubert~Lampeitl,\altaffilmark{5}
John~Marriner,\altaffilmark{14}
Ramon~Miquel,\altaffilmark{15,20}
Mercedes~Moll\'a,\altaffilmark{21}
Robert~C.~Nichol,\altaffilmark{5}
Jakob~Nordin,\altaffilmark{18}
Adam~G.~Riess,\altaffilmark{22,23}
Masao~Sako,\altaffilmark{24}
Donald~P.~Schneider,\altaffilmark{25}
Jesper~Sollerman,\altaffilmark{17,26}
J.~Craig~Wheeler,\altaffilmark{27}
Linda~\"{O}stman,\altaffilmark{18}
Dmitry~Bizyaev,\altaffilmark{28}
Howard~Brewington,\altaffilmark{28}
Elena~Malanushenko,\altaffilmark{28}
Viktor~Malanushenko,\altaffilmark{28}
Dan~Oravetz,\altaffilmark{28}
Kaike~Pan,\altaffilmark{28}
Audrey~Simmons,\altaffilmark{28}
and Stephanie~Snedden\altaffilmark{28}
}
\altaffiltext{1}{
Department of Physics and Astronomy, Rutgers, the State University 
of New Jersey, 136 Frelinghuysen Rd., Piscataway, NJ 08854.  
}
\altaffiltext{2}{
Department of Physics, University of Chicago, Chicago, IL 60637.
}
\altaffiltext{3}{
Kavli Institute for Cosmological Physics, The University of Chicago,
5640 South Ellis Ave., Chicago, IL 60637.
}
\altaffiltext{4}{
Department of Mathematics and Applied Mathematics, University of Cape
Town, Rondebosch 7701, South Africa.
}
\altaffiltext{5}{
Institute of Cosmology and Gravitation, Mercantile House, Hampshire 
Terrace, University of Portsmouth, Portsmouth PO1 2EG, United Kingdom.
}
\altaffiltext{6}{
South African Astronomical Observatory, P.O. Box 9, Observatory 7935, 
South Africa.
}
\altaffiltext{7}{
Department of Astronomy, University of Washington, Box 351580, Seattle, 
WA 98195.
}
\altaffiltext{8}{
Max Planck Institute for Extraterrestrial Physics, D-85748 Garching, 
Germany. 
}
\altaffiltext{9}{
Universitaets-Sternwarte Munich, 1 Scheinerstrasse, D-81679 Munich, Germany. 
}
\altaffiltext{10}{
Institut de Ci\`encies de l'Espai (IEEC-CSIC), Barcelona, Spain.
}
\altaffiltext{11}{
Department of Physics and Astronomy, Wayne State University, 
Detroit, MI 48202.          
}
\altaffiltext{12}{
Department of Astronomy, University of California, Berkeley, CA 
94720-3411. 
}
\altaffiltext{13}{
Department of Astronomy and Astrophysics, The University of Chicago, 
5640 South Ellis Avenue, Chicago, IL 60637.
}
\altaffiltext{14}{
Center for Astrophysics, Fermi National Accelerator Laboratory, P.O. 
Box 500, Batavia, IL 60510. 
}
\altaffiltext{15}{
Institut de F\'{\i}sica d'Altes Energies, Barcelona, Spain.
}
\altaffiltext{16}{
University of Notre Dame, 225 Nieuwland Science, Notre Dame, IN 46556-5670. 
}
\altaffiltext{17}{
The Oskar Klein Centre, Department of Astronomy,  Albanova, 
Stockholm University, SE-106 91 Stockholm, Sweden.
}
\altaffiltext{18}{
Department of Physics, Stockholm University, Albanova University 
Center, S-106 91 Stockholm, Sweden.
}
\altaffiltext{19}{
Institute of Astronomy, Graduate School of Science, University of 
Tokyo 2-21-1, Osawa, Mitaka, Tokyo 181-0015, Japan.
}
\altaffiltext{20}{
Instituci\'o Catalana de Recerca i Estudis Avan\c{c}ats, Barcelona,  Spain.
}
\altaffiltext{21}{
Centro de Investigaciones Energ\'eticas, Medioambientales y Tecnol\'{o}gicas, 
Madrid, Spain.
}
\altaffiltext{22}{
Space Telescope Science Institute, 3700 San Martin Drive, Baltimore, 
MD 21218.
}
\altaffiltext{23}{
Department of Physics and Astronomy, Johns Hopkins University, 3400 
North Charles Street, Baltimore, MD 21218.
}
\altaffiltext{24}{
Department of Physics and Astronomy, University of Pennsylvania, 
209 South 33rd Street, Philadelphia, PA 19104.
}
\altaffiltext{25}{
Department of Astronomy and Astrophysics, 525 Davey Laboratory, 
Pennsylvania State University, University Park, PA 16802.
}
\altaffiltext{26}{
Dark Cosmology Centre, Niels Bohr Institute, University of Copenhagen, 
Denmark.
}
\altaffiltext{27}{
Department of Astronomy, University of Texas, Austin, TX 78712.
}
\altaffiltext{28}{
Apache Point Observatory, P.O. Box 59, Sunspot, NM 88349.
}

\begin{abstract}
We present a measurement of the volumetric Type Ia 
supernova (SN~Ia) rate
based on data from the 
Sloan Digital Sky Survey II (SDSS-II) Supernova Survey. 
The adopted sample of supernovae (SNe)
includes \nsnezltohthree~SNe Ia at redshift $z \lesssim 0.3$, 
of which 
$\nsneconf~(\nsneconfpct\%)$  are spectroscopically identified as SNe~Ia. 
The remaining \nsnephoto\ SNe~Ia were identified through their light curves;
\nsnephotoz~of these objects have spectroscopic redshifts from
spectra of their host galaxy, and \nsnephotonoz\ have photometric redshifts
estimated from the SN light curves. Based on 
consideration of $87$ spectroscopically
confirmed non-Ia SNe discovered by the SDSS-II SN Survey, we 
estimate that 
$2.04^{+1.61}_{-0.95}\%$ 
of the photometric SNe~Ia may
be misidentified.
The sample of SNe Ia used in this measurement 
represents an order of magnitude increase
in the statistics for SN Ia rate measurements in the 
redshift range covered by the \sns. 
If we assume a SN Ia rate that is constant at 
low redshift ($z<0.15$), then the 
SN observations can be used to infer a value of the 
SN rate of
$r_V = (\lowzvalonefive^{+\lowzvalonefivehistat+\lowzvalonefivehisyst}_{-\lowzvalonefivelowstat-\lowzvalonefivelowsyst})$
$\times 10^{-5}$
$\mathrm{SNe}$ $\mathrm{yr}^{-1}$ $\mathrm{Mpc}^{-3}$
$(H_0/(70$~{km~s$^{-1}$~Mpc$^{-1}$}))$^{3}$
at a mean redshift of $\sim 0.12$, based on
79 SNe~Ia of which 72 are spectroscopically confirmed.
However, the large sample of SNe Ia included in this study allows us to 
place constraints on the redshift dependence of the 
SN~Ia rate based on 
the \sns~data alone. 
Fitting a power-law model of the SN rate evolution,
$r_{V}(z) = A_p \times ((1+z)/(1+z_{0}))^{\nu}$, 
over the redshift range $0.0 < z < 0.3$
with $z_{0} = \zoff$, results 
in 
$A_p = (\apval^{+\apvalhi}_{-\apvallo}) \times 10^{-5}$ 
$\mathrm{SNe}$ $\mathrm{yr}^{-1}$ $\mathrm{Mpc}^{-3}$
$(H_0/(70$~{km~s$^{-1}$~Mpc$^{-1}$}))$^{3}$
and
$\nu = \nuval^{+\nuvalhi}_{-\nuvallo}$.
\end{abstract}


\clearpage

\setcounter{footnote}{0}

\section{Introduction}
\label{sec:midzintro}

Type Ia supernovae (SNe~Ia) occupy a prominent position in 
contemporary astrophysics, in part due to their utility as 
cosmological distance indicators
\citep[for a review, see ][]{Filippenko_05}.
The observed correlation
between the peak luminosity and the rate of decline
for SNe~Ia \citep{Psk_77, Phillips_93}
has been exploited to improve the accuracy
of measured distances to SNe~Ia and thereby place important
constraints on cosmological 
models \citep[e.g.,][]{Riess_98, Perlmutter_99, Riess_04, Astier_06, 
Woodvasey_07, Riess_07, Hicken_09, Riess_09, Freedman_09, Kessler_09}. 
However, the SN~Ia 
decline rate vs. peak luminosity correlation is mainly an empirically 
determined phenomenon, and the exact nature of the 
progenitor systems that give rise to SNe~Ia
remains uncertain.
A better understanding of 
SN~Ia progenitor systems is desirable both for investigations
of fundamental astrophysics (e.g., binary star evolution
and explosion physics) and to provide a theoretical  
foundation for understanding any possible 
evolution of SN~Ia properties (such as the decline rate vs. peak 
luminosity correlation) with redshift that 
could cause additional systematic effects in distance 
measurements.

The SN~Ia rate can be used to place important constraints 
on the progenitor systems 
of SNe~Ia. In general, the SN~Ia rate can be expressed as a 
delay function
convolved with the cosmic star-formation rate (SFR)
\citep[e.g.,][]{Greggio_05}. 
That is,

\begin{equation}
\label{eqn:ddtrate}
r(t) = \int_{0}^{t}{k_{\Gamma}(t')~\Psi(t')~A_{\mathrm{SN}}(t-t')~D(t-t')~dt'},
\end{equation}

\noindent where $r(t)$ is the SN rate,
$\Psi(t')$ is the SFR, 
$k_{\Gamma}(t')$ is the number of stars per unit mass for the 
population formed at epoch $t'$, 
$A_{\mathrm{SN}}(t-t')$ is the number of stars from the 
population that will result in SN explosions, 
and $D(t-t')$ is a distribution of delay times 
between the formation of a stellar system and the 
resulting SN explosion.
The delay function varies depending on the model assumed 
for the progenitors of SNe~Ia, and measurements of the 
SN rate, in combination with measurements of 
the cosmic SFR, can therefore place 
observational constraints on SN~Ia progenitor models.
We emphasize that according to Eq. (\ref{eqn:ddtrate}), constraints on 
SN Ia progenitor systems rely not only on precise measurements of the 
SN rate, but also on measurements of the cosmic SFR.
At present, measurements of the cosmic SFR (as a function of time)
suffer from significant uncertainties, thus
complicating the interpretation of the cosmic SN rate
in terms of delay functions \citep{Forster_06}.
In this paper, we focus on presenting our SN rate 
measurements and will not pursue detailed comparisons to 
the cosmic SFR.

The SN rate was first measured by \citet{Zwicky_38}, who found it 
to be approximately ``one SN per few hundred years per average 
nebula,'' in the local universe.
Subsequently, improvements in astronomical technology as well
as increased observing time dedicated to SN searches 
have led to more precise SN rate measurements, spanning
a wide range of redshifts. 
In the local universe, the SN~Ia rate 
has been measured by \citet{Cappellaro_99} from 
$\sim 140$ SNe~Ia and by \citet{Li_10a, Li_10b} from $\sim 930$ SNe~Ia.  
At intermediate redshifts ($0.1 \lesssim z \lesssim 0.5$), the 
SN~Ia rate has been measured by many authors \citep[e.g.,][]{
Hardin_00,
Pain_02,
Madgwick_03,
Tonry_03,
Blanc_04,
Neill_06,
Sullivan_06,
Botticella_08}.
At high redshifts, the SN~Ia rate has been measured with data from the 
{\it Hubble Space Telescope (HST)} by \citet{Dahlen_04, Dahlen_08}.
All of these SN rate measurements are based
on SN~Ia samples that are primarily spectroscopically identified 
and were determined in a manner similar to that of the
\sns\ SN rate analysis presented here. 
In addition to these measurements, a number
of authors have presented SN~Ia rate analyses based on photometric 
identification of SNe~Ia, in many cases with only a few photometric observations.
These include measurement of 
the intermediate-redshift rate
by \citet{Horesh_08},
the intermediate-to-high redshift rate by
\citet{Barris_06},
and the high-redshift rate 
by \citet{Poznanski_07a} 
and
\citet{Kuznetsova_08}.

A precise measurement of the 
low-redshift ($z < 0.12$) SN~Ia rate, based on 17 SNe Ia from the 
first season of the \sns, was given by \citet{Dilday_08a}.
In the present paper we discuss an extension of this volumetric 
SN Ia rate measurement to a higher redshift limit, based on
all three seasons of the \sns\ \citep{Frieman_08}.
Including SNe from three years of the \sns\ and 
considering a larger redshift range results in a 
major increase in the number of SNe used for the 
rate measurement. At low redshifts, the SN rate measurements
discussed here have the same high purity and completeness as for the 
low-redshift rate from the first season discussed by \citet{Dilday_08a},
but with increased statistical power.
Inclusion of higher-redshift SNe allows for investigation of the 
redshift dependence of the SN~Ia rate over the range covered by the 
\sns. 
However, at higher redshifts, systematic uncertainties become 
increasingly important and eventually dominate the error 
budget.
The efficiency studies and SN selection functions described herein
have also been used to estimate the SN~Ia rate as an explicit function
of the properties of their host galaxies \citep{Smith_10} and for studies of the 
SN~Ia rate in clusters of galaxies \citep{Dilday_10}.

The rest of this paper is organized as follows. In \S \ref{sec:sdsssnobs}
we briefly describe the observations and SN search strategy of the \sns.
Section \ref{sec:midzsample} discusses selection of the 
SN rate sample from the \sns~data, and
\S \ref{sec:midzeffs} determines the 
efficiency for SN selection. 
We present our measurement of the SN~Ia rate in 
\S \ref{sec:results}, and our conclusions
are summarized in \S \ref{sec:midzconclude}. 
Whenever necessary, we assume a flat $\Lambda$CDM universe
with $\Omega_m = 0.3$, $\Omega_{\Lambda} = 0.7$, and 
$H_0 = 70~\mathrm{km}~\mathrm{s}^{-1}~\mathrm{Mpc}^{-1}$.

\section{SDSS-II Supernova Survey Observations}
\label{sec:sdsssnobs}

Here we briefly describe aspects of the {\sns} most
relevant to the present SN rate analysis. 
Much of the material in this section is also relevant to the
SN rate studies described by \citet{Dilday_10},
and is discussed more fully therein.
The survey is described in more
detail by \citet{Frieman_08}, and the SN detection algorithms
are given by \citet{Sako_08}. Additional details
of the survey observations and 
the use of {\it in situ} artificial SNe 
for determining SN detection efficiencies 
are discussed by
\citet{Dilday_08a}.
A technical summary of the SDSS is given by~\citet{York_00}.
Details of the survey calibration 
are provided by \citet{Hogg_01}, \citet{Smith_02}, and \citet{Tucker_06}. 
The data processing and quality assessment are discussed by~\citet{Ivezic_04}, 
and the photometric pipeline is described by~\citet{Lupton_99}.

The \sns~was carried out during 
the Fall (September--November) of 2005--2007, 
using the 2.5~m telescope \citep{SDSS_telescope} at 
Apache Point Observatory (Sacramento Peak, New Mexico).
Observations were obtained 
in the SDSS $ugriz$ filters \citep{Fukugita_96}
with a wide-field CCD camera \citep{Gunn_98},
operating in time-delay-and-integrate (TDI, or drift scan) mode. 
The region of the sky covered by the 
\sns~(designated Stripe 82; see \citealt{SDSS_EDR})
was bounded by 
$-60^{\circ} < \alpha_{\rm J2000} < 60^{\circ}$ and 
$-1.258^{\circ} < \delta_{\rm J2000} < 1.258^{\circ}$.
On average, any given part of this $\sim 300$ square degree 
area was imaged once every 4 days during the survey operations.

Difference images were produced in the SDSS $gri$ filter bands
by subtracting 
template images, constructed from 
previous survey observations of the region, 
using an implementation of the methods described 
by~\citet{Alard_98}.  
The difference images were searched for positive fluctuations
using the {\tt DoPHOT} photometry 
and object detection package~\citep{Schechter_93};
typical limits ($10 \sigma$ above background) 
for the \sns\ were
$g \approx 21.8$, $r \approx 21.5$, and $i \approx 21.2$ mag.
A combination of software cuts and human visual inspection was then used to 
identify promising SN candidates from the full set of transient detections.
As a key component of prioritizing 
SN candidates for follow-up spectroscopic observations, 
the light curves of SN candidates were fit to 
models of Type Ia, Type Ib/c, and Type II
SNe (see \citealt{Filippenko_97} for a review of SN types).
This procedure is referred to
as ``photometric typing,'' and is described in detail 
by \citet{Sako_08}.

Spectroscopic observations
for both SN classification and redshift determination were provided
by a number of different telescopes. The spectra of the SNe utilized in the 
present SN rate analysis were provided by 
the Hobby-Eberly 9.2~m at McDonald Obseratory, 
the Astrophysical Research Consortium 3.5~m at Apache Point
Observatory, 
the Hiltner 2.4~m at the Michigan-Dartmouth-MIT Observatory, 
the Subaru 8.2~m at the National Astronomical Observatory
of Japan, the Keck-I 10~m at the W. M. Keck Observatory, 
the Mayall 3.8~m at Kitt Peak National Observatory,
the 3.5~m ESO New Technology Telescope (NTT) at the European
Southern Observatory,
the SALT 11~m (9.5~m clear aperture) at the South African Astronomical Observatory,
and the
2.6~m Nordic Optical Telescope, 3.5~m Telescopio  
Nazionale Galileo, and 4.2~m William Herschel Telescope at the
Observatorio del Roque de los Muchachos.
Details of the \sns\ spectroscopic data reductions are given by \citet{Zheng_08}.
Comparison to high-quality SDSS galaxy spectra shows that SN spectroscopic redshifts 
are accurate to $\sim 0.0005$ when galaxy emission features are used 
and to $\sim 0.005$ when SN features are used. In either case, the uncertainties in the 
spectroscopic SN redshifts are negligible for the SN-rate studies considered here.

While the difference imaging pipeline used during the SN search
provides initial photometric measurements, subsequent to the search more 
precise SN photometry is provided 
using a scene modeling photometry (SMP) technique developed 
by~\citet{Holtzman_08}. 
The final analysis of SN light curves and the selection 
cuts used to define the SN rate sample 
discussed in this paper are based on SMP.

\section{SN Ia Sample for the Rate Measurement}
\label{sec:midzsample}

\subsection{SN Selection Requirements}
\label{sec:midzsnsection}

For the \sns~measurement of the low-redshift 
SN~Ia rate \citep{Dilday_08a},
we included in the SN Ia sample all spectroscopically 
confirmed SNe Ia at $z < 0.12$, subject to a set of objective selection 
criteria that can be robustly modeled with our 
SN Monte Carlo (MC) simulations.
To account for spectroscopic incompleteness, we used the 
MLCS2k2 SN Ia model \citep{Jha_07} to analyze the
SMP (\S \ref{sec:sdsssnobs}) light curves for a set 
of $\sim 1000$ photometric SN candidates, 
which comprised $\sim 500$
``best'' SN~Ia candidates and $\sim 500$ randomly chosen SN candidates.

In the present analysis, we adopt a somewhat different approach
to selecting the SN sample for use in measurement of the SN rate.
Rather than focusing on the low-redshift ($z < 0.12$) SNe, 
which can be demonstrated to be a complete sample, 
we define objective selection criteria for 
SNe Ia at all redshifts, and determine the completeness
of the resulting samples based on analysis of simulated samples
of SNe. 
As discussed in \S \ref{sec:sdsssnobs}, during the survey the 
search-photometry light curves of SN candidates were fit to models
of Type Ia, Type Ib/c, and Type II SNe, and the results were used as a factor
in prioritizing our spectroscopic follow-up resources.

In addition, as a method of searching for photometric SNe~Ia 
subsequent to the survey, 
the search-photometry light curves were used to define
a Bayesian probability for each SN candidate to be a SN of a given type.
This was done by marginalizing over the light-curve fit parameters to
obtain the {\it Bayesian evidence} and requiring that 
the evidence for the three SN types sums to 1. This defines the 
``probability,'' $p_{\mathrm{T}}$, for an object to 
be a SN of type T.
This quantity can be considered a probability in the sense that it is bounded 
by $ 0 < p_{\mathrm{T}} < 1$, and is normalized to 1 ($\Sigma_{\mathrm{T}} 
\, p_{\mathrm{T}} = 1$).
However, this procedure makes the initial assumption that the object is a
SN (i.e., that the three types T = Ia, Ib/c, II are 
exhaustive), and does not allow for other possibilities for the
identity of the object (e.g., active galactic nucleus). Despite
this caveat, the quantities $p_{\mathrm{T}}$ are 
useful statistics for analyzing the search photometry light curves.
The procedure is motivated by, and modeled after, that discussed by
\citet{Kuznetsova_07} and \citet{Poznanski_07}.

The selection criteria for SN candidates that we impose on the 
photometric-typing fits (\S \ref{sec:sdsssnobs}) 
to the search photometry light curves
are as follows:

\begin{enumerate}

\item Bayesian $p_\mathrm{Ia} > 0.45$.
\item At least three search-discovery epochs.
\item  If the candidate has more than five search-photometry epochs, 
the best-fit SN~Ia model is not SN 2005gj.

\end{enumerate}

\noindent 
These selection criteria were determined by correlating the fit results from the full 
analysis of the SMP light curves for the $\sim 1000$ photometric SN candidates from the 
2005 season with statistics of the corresponding photometric-typing fits to the search 
photometry, and looking for a combination of cuts that would result in a sample of SN 
candidates with high purity and completeness with respect to SNe Ia.
Several possible statistics of the photometric-typing fits were considered to 
see whether they would improve the
efficiency for selecting SNe Ia from the search-photometry SN candidates. 
The conclusion of these
correlation studies was that the Bayesian probability, $p_{\mathrm{Ia}}$, 
is the best single quantity to 
consider for selecting a large fraction of SNe Ia, and no significant improvement was
found by considering additional fit statistics, such 
as the value of the reduced $\chi^{2}$ statistic for the fit.

The peculiar SN Ia 2005gj, which has 
a flat light curve after maximum brightness \citep{Aldering_06,Prieto_07},
is included as one of the SN Ia 
light-curve models in the photometric-typing fits.
The requirement that the best-fit SN {\it not} 
be SN 2005gj is effectively intended to remove both peculiar SNe Ia and 
other non-SN transients, such as active galactic nuclei.
Some core-collapse SNe are well fit by the broad light curve of 
SN 2005gj, and this cut also serves as a way for rejecting these from our 
SN Ia sample selection.
Search-discovery epochs refer to epochs for which the transient object
was detected by the survey difference imaging and object detection 
pipeline \citep{Sako_08}.

This sample selection, based on the photometric-typing procedure, resulted in 
$\sim 600$ SN Ia candidates per observing season of the \sns.
Scene modeling photometry was then generated for these candidates, 
producing more reliable photometry
and providing measurements at additional observing epochs, compared
with the SN search photometry.
In addition to the requirements on the photometric-typing fits, 
we require the SMP light curves for the candidates to satisfy similar 
selection criteria on light-curve sampling and fit quality 
to those discussed by \citet{Dilday_08a}. We list these criteria
below:

\begin{enumerate}

\item $-51^{\circ} < \alpha_{\rm J2000} < 57^{\circ}$.

\item There are photometric observations on at least five separate 
epochs between $-20$ days and $+60$ days 
relative to $B$-band peak light in the SN rest frame.

\item At least one epoch with signal-to-noise ratio $> 5$ in each of 
$g$, $r$, and $i$ (not necessarily the same epoch in each passband). 

\item At least one photometric observation at least two days 
{\it before} $B$-band peak light in the SN rest frame.

\item At least one photometric observation at least ten days 
{\it after} $B$-band peak light in the SN rest frame.

\item MLCS2k2 light-curve fit probability $> 0.001$.

\item MLCS2k2 light-curve fit parameter $\Delta > -0.4$.

\end{enumerate}

\noindent The first requirement states that the SN is
within the right-ascension range of the calibration-star catalog. The second and
third requirements ensure that the object is a significant 
and authentic astrophysical transient. The fourth and fifth 
requirements are imposed so that we have a robust measurement of the time of 
maximum brightness for the SN candidates as well as a reliable 
measurement of the light-curve decline, which is 
useful for rejecting Type II SNe. 
The sixth requirement is used to reject peculiar SNe~Ia
that are not well represented by the MLCS2k2
light-curve model,
as well as non-SN~Ia transient objects.
The seventh requirement is additionally used to reject objects with 
flat light curves such as SNe~II and active galactic nuclei. 
The low-redshift SN data used to define the MLCS2k2
model only exhibit values of the light-curve shape parameter
$\Delta \gtrsim -0.35$, so a cut at $\Delta > -0.4$ 
specifies that the object is within the valid range of the
MLCS2k2 model, with some allowance for measurement error.

\subsection{SN Sample}
\label{sec:midzsnsample}

\begin{figure} [!t] 
\begin{center}
\includegraphics[width=5.75in]{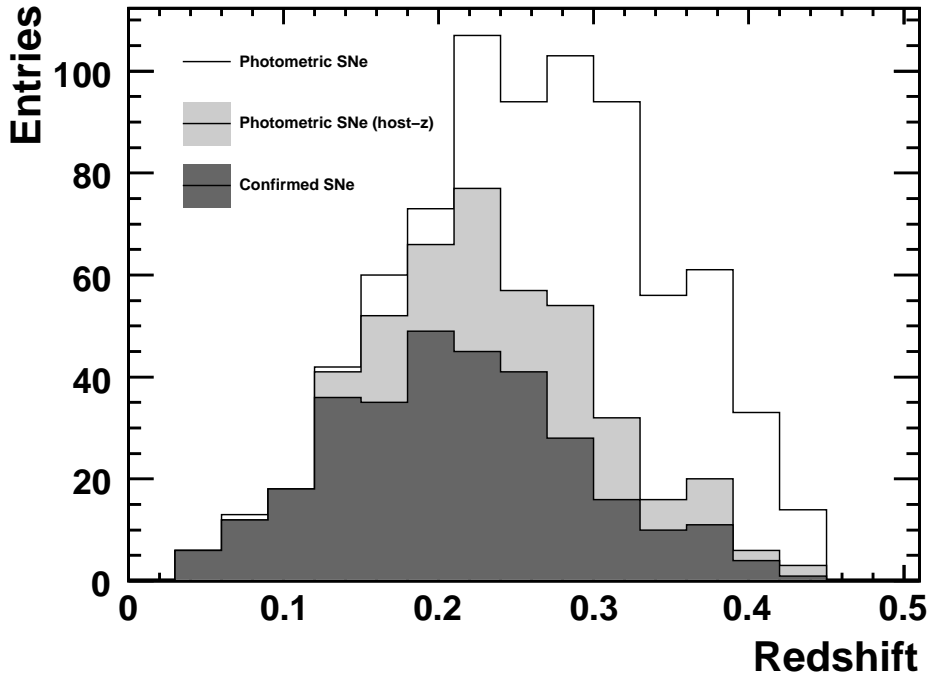}
\end{center}
\caption{
The redshift distribution for the \nsnetotall~\sns~SNe {\it passing} 
all selection criteria. The 
dark gray,
light gray, and 
white
shading
represent spectroscopically confirmed SNe Ia, photometric 
SNe {\it with} measured host-galaxy 
redshifts, and photometric SNe {\it without} measured host-galaxy 
redshifts, respectively.
 }
\label{fig:fmidz1}
\end{figure}

Over the entire redshift range of SNe discovered by 
the \sns~($ z \lesssim 0.45$), there 
are \nsnetotall~SN Ia candidates 
(\nsnetotallconf~spectroscopically confirmed)
that satisfy the selection 
criteria above. The redshift distribution for these SNe is
shown in Figure \ref{fig:fmidz1}. However, as will be 
discussed in \S \ref{sec:midzeffs},
the systematic uncertainty related to our sample selection 
becomes dominant for $ z \gtrsim 0.2$, 
thereby reducing our ability for making precise SN Ia rate 
measurements.
The numbers
of SNe for several values of the maximum redshift
are given in Table \ref{tab:midzsncount}.
Spectroscopically confirmed SNe Ia from this sample are listed in Table \ref{tab:snerates2_conf}.

\begin{deluxetable}{crrrr}
\tablecolumns{5}
\tabletypesize{\small}
\singlespace
\tablewidth{0pc}
\tablecaption{Number of SNe Ia for Rate Measurement
\label{tab:midzsncount}
}
\tablehead{
\colhead{Redshift} & 
\colhead{Confirmed} &    
\colhead{Photometric} &
\colhead{Photometric} & 
\colhead{Total} \\ 
 
\colhead{Limit} &
\colhead{} & 
\colhead{(Spect-$z$)} & 
\colhead{(Photo-$z$)} & 
\colhead{} 
}

\renewcommand{\arraystretch}{1.5}
\startdata
$0.15$   &  72 $(91 \%$) &   5 $( 6 \%$)  &   2  $( 3 \%$)  &  79 \\ 
$0.20$   & 140 $(74 \%$) &  35 $(18 \%$)  &  15  $( 8 \%$)  & 190 \\ 
$0.25$   & 217 $(62 \%$) &  76 $(22 \%$)  &  57  $(16 \%$)  & 350 \\ 
$0.30$   & 270 $(52 \%$) & 113 $(22 \%$)  & 133  $(26 \%$)  & 516 \\ 
$\infty$ & 312 $(40 \%$) & 148 $(19 \%$)  & 314  $(41 \%$)  & 774 \\ 
\enddata


\end{deluxetable}

SNe for which we have photometric observations of the light curve,
but do not have any spectroscopic observations to determine 
the spectral type of the SN, are referred to as {\it photometric SNe}.
Photometric SNe fall into two classes: 
(i) those that have a precisely (i.e., spectroscopically) 
measured redshift for their host galaxy, and
(ii) those that do not have a precisely measured redshift for 
their host galaxy.
When the redshift for a photometric SN candidate is unknown,
the candidate light curve is analyzed with the ``photo-$z$''
option in the flux-based MLCS2k2 light-curve fitter \citep{Dilday_08a}.
To fit SN light curves for redshift, we assume
a cosmological model, and hence a distance-vs.-redshift relation,
in order to take advantage of knowledge of the absolute
magnitude of SNe~Ia. In addition to redshift, the 
SN light curves are fit for the time of maximum 
(in the SN rest-frame $B$ band),
the luminosity parameter $\Delta$, and the extinction parameter
$A_V$; see \citet{Jha_07} for a comprehensive discussion of MLCS2k2.

\begin{figure} [!t] 
\begin{center}
\includegraphics[width=5.75in]{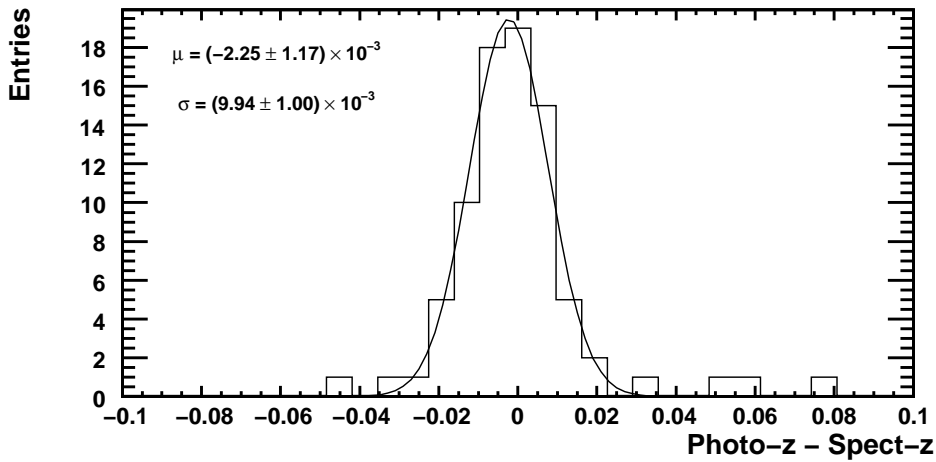}
\end{center}
\caption{
Redshift residuals (photo-$z$ minus spectroscopic-$z$)
for photo-$z$ fits to a sample of 
spectroscopically confirmed SNe Ia and 
photometric SNe Ia {\it with} spectroscopic redshifts
at $z < 0.15$. 
The best-fit Gaussian function is overlayed. Here,
$\mu$ refers to the mean and $\sigma$ to the standard deviation
of the best-fit function.
This plot illustrates that the 
SN photo-$z$ fits are negligibly biased, and accurate to 
$\sim 0.01$ at low redshift.
}
\label{fig:fpzresid}
\end{figure}

To investigate the accuracy and precision of the photo-$z$ fits,
we carry out photo-$z$ fits to the spectroscopically confirmed SNe Ia, as well as to 
the photometric SNe Ia {\it with} spectroscopically measured redshifts of
$z < 0.15$. At such low redshifts, this is essentially 
a complete sample of SNe Ia.
A plot of the residuals for the photometric redshifts is
shown in Figure \ref{fig:fpzresid}, 
illustrating that the 
SN photo-$z$ fits are negligibly biased, and accurate to 
$\sim 0.01$ at low redshift.
The numbers of SNe from categories (i) and (ii) that satisfy the
selection criteria are given, for 
several values of limiting redshift, in Table \ref{tab:midzsncount}.
Photometric SNe from categories (i) and (ii) are listed in Table
\ref{tab:snerates2_105} and Table \ref{tab:snerates2_photoz}, respectively. 

\subsection{Bias Correction for the Observed Redshift Distribution}
\label{sec:dndzbias}
\begin{figure} [!t] 
\begin{center}
\includegraphics[width=5.75in]{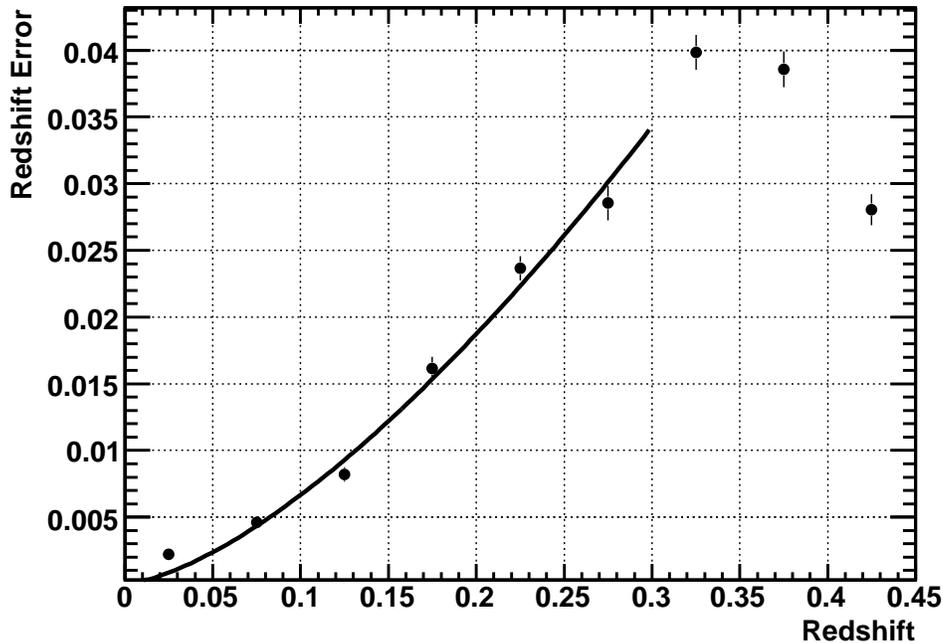}
\end{center}
\caption{
The standard deviation of the SN photo-$z$ estimates as a function of fitted photo-$z$. 
The points show the mean value of the photo-$z$ error, in bins of 
width $\delta z = 0.05$. The error bars 
represent the uncertainty in the mean, and the solid line represents the best-fit 
power law for the interval $0<z<0.3$.
 }
\label{fig:dpzvspz}
\end{figure}

The observed redshift distributions for the SNe~Ia
from the \sns~that satisfy the selection 
criteria of \S \ref{sec:midzsnsection} are shown in Figure \ref{fig:fmidz1}. 
As the number distribution
is not constant with redshift, 
nonzero measurement error of the redshifts will
result in a bias in the measured number distribution.
In general, the observed number distribution as a function of 
redshift, 
$\eta(z)$, 
is related to the true 
redshift distribution,
$\eta_{0}(z)$, 
through

\begin{equation} 
\eta(z) = 
\int_{-\infty}^{\infty} \eta_{0}(z') ~p(z|z')~ dz',
\end{equation} 

\noindent where $p(z|z')$ is the probability that a SN at redshift $z'$ will
have a measured redshift of $z$. 
The number distribution, $\eta_{0}(z)$, is related to the volumetric rate,
$r_{V}(z)$, and the redshift dependent efficiency, $\epsilon(z)$, through

\begin{equation} 
\eta_{0}(z)~dz \propto
\frac{r_{V}(z)~\epsilon(z)}{1+z}~\frac{dV}{dz}~dz,
\end{equation} 

\noindent where $dV/dz$ is the volume element at redshift $z$.
In what follows, we use a Gaussian approximation for 
$p(z|z')$, which in most cases is a good representation of 
the SDSS-SN photo-$z$ errors,

\begin{equation}
p(z|z') = \frac{1}{\sqrt{2 \pi} \sigma(z')}~e^{-(z-z')^{2}/2 \sigma^2(z')},
\end{equation}

\noindent where $\sigma(z')$ is the standard deviation of the SN photo-$z$.
To determine $\sigma(z')$, we perform photo-$z$ fits for 
all SNe that pass our selection criteria and then fit 
a power law for the typical error in the photo-$z$. 
The photo-$z$ error as a function of 
fitted photo-$z$ is shown in Figure \ref{fig:dpzvspz}, along with the
best-fit power law, $\sigma(z) = A \, z^{k}$. The best fit has
$A \approx 0.2$ and $k \approx 1.5$. 

\begin{figure} [!t] 
\begin{center}
\includegraphics[width=5.75in]{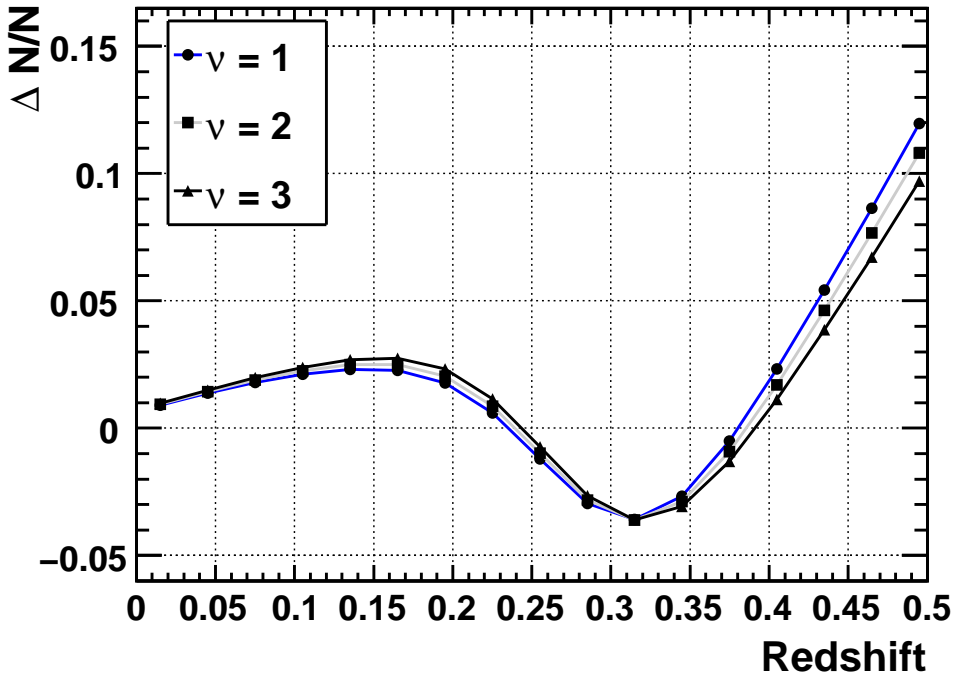}
\end{center}
\caption{
The bias, $\Delta N/N$, in an observed distribution of SNe, 
with a typical error in the 
measured redshifts (\S \ref{sec:dndzbias}) and 
discovery efficiency (\S \ref{sec:midzsnsection}) as determined for the 
\sns. The bias is plotted for a range of power-law SN rate models, 
$r(z) = A_p \, (1+z)^{\nu}$.
}
\label{fig:dndzbias}
\end{figure}

To estimate the bias in the observed SN number distribution, we 
integrate both $\eta(z)$ and $\eta_{0}(z)$ for a range of SN rate models,
$r(z) \propto (1+z)^{\nu}$, over the redshift bins shown in Figure \ref{fig:fmidz1}.
The resulting bias, defined as $\Delta N/N = (N - N_{0})/N$, where
$N$ is the number of observed SNe and $N_{0}$ is the 
number of underlying SNe in each bin, is
at the few percent level and is shown in Figure \ref{fig:dndzbias}.
We note that we have considered here a hypothetical SN sample where all the 
redshifts are determined photometrically. The bias 
in the observed \sns~SN distribution will be 
much smaller, since many of the redshifts are
determined spectroscopically. 
In Table \ref{tab:midzratez1} we list the 
bias correction 
appropriate for our best-fit power-law SN rate model,
$r_V(z) \propto (1+z)^{\nuval}$,
computed over bins in redshift of width $\delta z = 0.05$.
The bias due to the use of SN photometric redshifts 
is negligible in comparison to the statistical and 
systematic uncertainties on the SN Ia rate measurements.
Additionally, we note that in \S \ref{sec:ratevsz} we  
fit the \sns~data to models of the SN rate
using an unbinned maximum likelihood that 
properly accounts for the bias discussed here.

\subsection{Estimating Contamination from Non-Ia SNe}
\label{sec:contamination}

In studies of SNe~Ia that include photometrically
identified SNe, one must correct for contamination
from non-Ia (mainly core-collapse) SNe and evaluate
the corresponding systematic uncertainty.
Non-Ia SNe form a less
homogeneous set than Type Ia SNe and, in contrast to SNe Ia,  
no general parametric models exist to describe the 
light curves of non-Ia SNe. 
In addition, owing to the strong interest in the use of SNe~Ia
as cosmological distance indicators, there is often an explicit observational bias 
{\it against} spectroscopic follow-up observations of non-Ia SNe in modern SN surveys 
\citep[e.g.,][]{Sullivan_06a,Sako_08}. As the global set
of well-observed SNe Ia has grown, this limitation for future SN~Ia
studies has been recognized, and non-Ia SN 
observations, in part to  better characterize the 
underlying SN population, are gaining increased attention 
\citep[e.g.,][]{Galyam_05,Galyam_07}.

\begin{figure} [!t] 
\begin{center}
\includegraphics[width=5.75in]{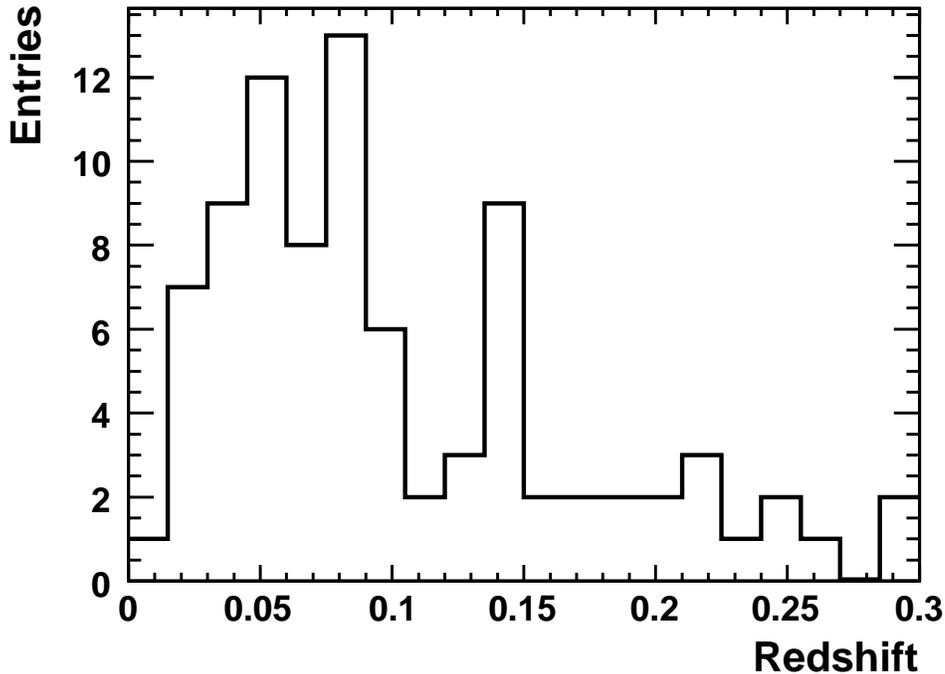}
\end{center}
\caption{Redshift distribution for spectroscopically confirmed non-Ia SNe 
for the 2005--2007 observing seasons of the \sns.
 }
\label{fig:noniaz}
\end{figure}

For these reasons, it is difficult to quantify contamination 
of the set of photometric SN~Ia candidates from non-Ia SNe in the same way 
that we have treated determination of the completeness 
of the SN Ia sample --- namely,  by modeling the SN survey observations of the
underlying population with our SN Monte Carlo simulations.
To place limits on the expected level of contamination of the photometric SN
sample by non-Ia SNe, we instead consider the set of 
spectroscopically confirmed non-Ia SNe 
from the \sns. 
There are 
\nnoniaonetwo\ spectroscopically confirmed non-Ia SNe from the first
two years of the \sns, 
and an additional 
\nnoniathree\ from the third year. 
The reason for the greater relative number of non-Ia SNe in 
the third year is that, in addition to the usual SN~Ia search, 
the third season included spectroscopic observing time on the Subaru telescope
specifically allocated for Type II SNe \citep{Dandrea_10}.
The redshift distribution for the 
spectroscopically confirmed non-Ia SNe from all 
three years of the \sns\ is 
shown in Figure \ref{fig:noniaz}.

\begin{figure} [!t] 
\begin{center}
\includegraphics[width=5.75in]{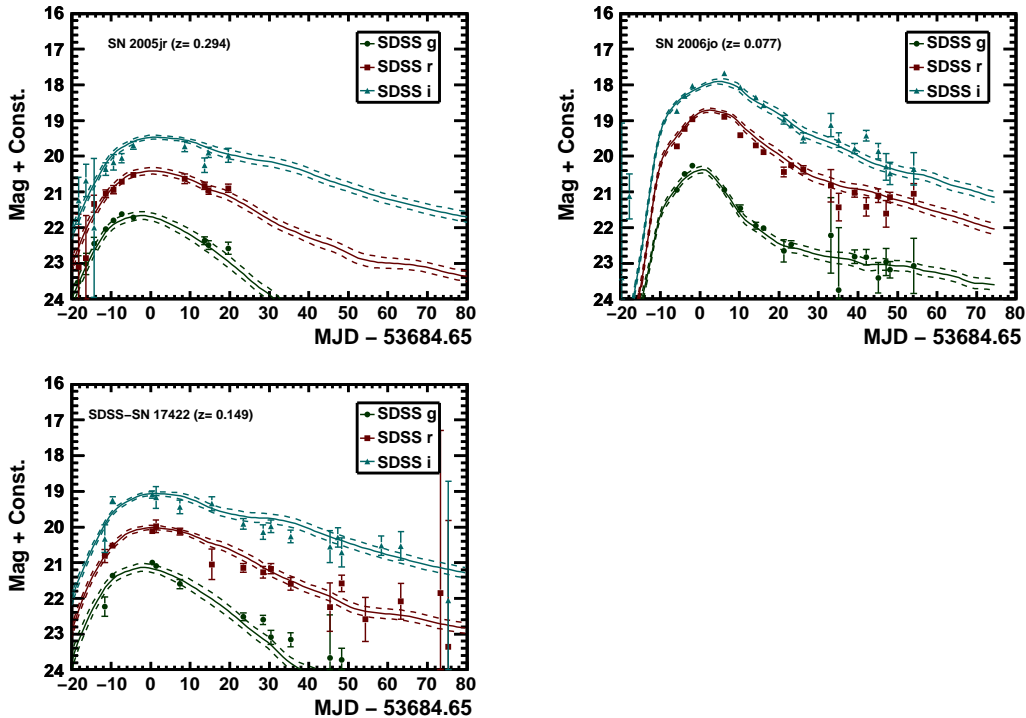}
\end{center}
\caption{
Photo-$z$ light-curve fits, to a SN~Ia model, for the non-Ia SNe 
SDSS-SN 8679 (top-left panel),
SDSS-SN 14492 (top-right panel),
and SDSS-SN 17422 (bottom-right panel).
The points represent the observed SN magnitudes, as 
a function of time, in the observer frame.
The solid lines
represent the best-fitting model light curves, for the SDSS 
$g$, $r$, and $i$ filter bands, and the dashed lines represent the
corresponding 1$\sigma$ MLCS2k2 model errors.
For clarity, the $g$, $r$, and $i$ light curves 
are offset by +0, +1, and +2 mag,
respectively. 
}
\label{fig:nonialcs}
\end{figure}

To investigate possible contamination from non-Ia SNe, 
we apply to the set of non-Ia SNe the same 
light-curve fits (to a SN Ia model) and selection criteria
that are used in defining the SN Ia sample.  
As the set of photometric SN~Ia candidates includes 
SNe both with and without spectroscopically measured redshifts,  
we consider contamination from both 
distance and photo-$z$ fits for the confirmed non-Ia SN sample.
If we fix the redshift to its spectroscopically determined 
value and fit the set of non-Ia SNe 
for distance, 
two events satisfy the selection criteria for the
SN Ia rate sample. 
These are SDSS-SN 14492 (SN 2006jo), a SN~Ib
at $z = 0.077$, and SDSS-SN 17422 (no IAU designation), a
SN~II at $z = 0.149$. 
The light-curve fits using the MLCS2k2 SN Ia model for these
SNe are shown in Figure \ref{fig:nonialcs}. 
If we instead fit the set of non-Ia SNe for photo-$z$,
three of them satisfy the selection criteria for the
SN Ia rate sample. These include the two non-Ia SNe 
mentioned above, as well as
SDSS-SN 8679 (SN 2005jr), a SN~IIn at $z=0.294$.
If we assume that the fraction of  
non-Ia SNe that satisfy the selection criteria, $q_{\mathrm{CC}}$, is a random variable that follows 
a binomial distribution, then an observation of two successful
events 
out of \nnoniaall\ in total 
gives 1$\sigma$ limits on $q_{\mathrm{CC}}$ of 
$0.023^{+0.030}_{-0.015}$. 
An observation of three successful events
gives 1$\sigma$ limits on $q_{\mathrm{CC}}$ of 
$0.035^{+0.032}_{-0.019}$. 

Some care must be taken when interpreting the fits to the spectroscopically 
confirmed non-Ia SNe as an estimate
of the false-positive rate.
The spectroscopic incompleteness of the \sns~for non-Ia SNe is not well constrained, and 
the set of spectroscopically confirmed non-Ia SNe is manifestly {\it not}
complete. As mentioned above, the distribution of non-Ia SN light-curve properties
is not well measured, so it is difficult to evaluate whether these 
non-Ia SNe are a representative sampling of the underlying population. 
However, 
as the \sns~has a built-in selection bias {\it against} non-Ia SNe \citep{Sako_08},
it can plausibly be claimed that any bias in the \sns~non-Ia sample is a bias
{\it toward} the most ``SN~Ia-like'' non-Ia SNe. With that being the case, our estimate
of $\sim 3\%$ 
probability for non-Ia SNe to satisfy our selection criteria 
can be considered a conservative upper bound.

To turn this into an estimate of the contamination of the photometric
SN sample from non-Ia SNe, we must also make an estimate of the 
non-Ia SN rate. In the redshift range containing the majority
of photometric SNe from the \sns, $ 0.2 < z < 0.3$,
the ratio of the non-Ia SN rate to the SN Ia rate 
has been measured by \citet{Botticella_08} 
as $(r_{\rm CC}/r_{\rm Ia})_{z=0.25} = 5.6 \pm 3.5$ 
and by \citet{Bazin_09}
as $(r_{\rm CC}/r_{\rm Ia})_{z=0.30} = 4.5 \pm 1.0$. 
Assuming the ratio is constant for $ 0.2 < z < 0.3$ and 
combining the two measurements gives a ratio
of the non-Ia SN rate to the SN Ia rate of
$r_{\rm CC}/r_{\rm Ia} = 4.6 \pm 1.0$. 

Furthermore, the above estimate of $\sim 3\%$ 
of non-Ia SNe satisfying the SN Ia selection
criteria was determined for SNe that were 
detected by the \sns\ and some estimate must be made 
of the detection efficiency.
To estimate the ratio of detection efficiency
for non-Ia SNe vs. SNe Ia, we employ the 
following procedure. The efficiency for 
SNe Ia to satisfy the MLCS2k2 component of the
selection function (i.e., the items listed in 
\S \ref{sec:midzsnsection})
as a function of redshift, 
shown in Figure \ref{fig:effz1}, is transformed
to an efficiency as a function of observer-frame peak 
magnitude by assuming a typical peak absolute 
magnitude for SNe~Ia of $M_B = -19.3$, and a distance 
modulus derived from a standard $\Lambda$CDM 
cosmological model. 
This is a good operational definition for detection, since
the primary requirements of the 
MLCS2k2 selection criteria are requirements on the SN sampling
and signal-to-noise ratios.
Although non-Ia SNe clearly differ from SNe Ia
in properties such as light-curve shapes and K-corrections, 
we will assume that the detection efficiency 
for non-Ia SNe can be described by the 
same function of observer-frame magnitude; the efficiency
function can then be mapped back to an efficiency as a function
of redshift, given an assumed absolute magnitude.

\citet{Richardson_02, Richardson_06} give estimates of the typical peak 
absolute magnitude 
for SNe Ib/c as $M_B = -18.07$,
for SNe II-P as $M_B = -16.98$,
and 
for SNe II-L as $M_B = -18.17$.
Assuming an absolute magnitude of $M_B \approx -18.0$ for 
non-Ia SNe, the ratio of detection efficiencies for
non-Ia SNe vs.~SNe Ia as a function of
redshift is then computed, and
is well approximated by a function
$\epsilon_{\rm CC}^{D}/\epsilon_{\rm Ia}^{D}
= 1/(1+e^{(z - z_0)/s_z})$.
The best-fit parameters are found to be 
$z_{0} = (0.204, ~0.213, ~0.211)$
and 
$s_{z} = (0.032, ~0.032, ~0.031)$
for the 2005, 2006, and 2007 seasons, respectively.
The ratio of non-Ia to Ia SNe in the set of photometric
SNe can then be estimated as 

\begin{equation}
\label{eqn:ccfrac}
\frac{N_{\rm CC}}{N_{\rm Ia}} = \frac{r_{\rm CC}}{r_{\rm Ia}}
~\frac{\epsilon^{q}_{\rm CC}}{\epsilon^{q}_{\rm Ia}}
~\frac{\epsilon^{D}_{\rm CC}}{\epsilon^{D}_{\rm Ia}},
\end{equation}

\noindent where $r$ is the SN rate, 
$\epsilon^{q}$ is the efficiency for a SN to satisfy the selection criteria
on light-curve shape and fit probability, and 
$\epsilon^{D}$ is the efficiency for detection, as described 
above.
We note that the ratio of core-collapse SNe to SNe Ia given by Eq.~\ref{eqn:ccfrac} 
is a function of redshift.
For the redshift range of interest, we take $r_{\rm CC}/r_{\rm Ia} = 4.6 \pm 1.0$
as discussed above.
The quantity 
$\epsilon^{q}_{\rm CC}/\epsilon^{q}_{\rm Ia}$ is 
$0.023^{+0.030}_{-0.015}$ 
for SNe with spectroscopically measured host-galaxy redshifts
and  
$0.035^{+0.032}_{-0.019}$
for SNe without spectroscopically measured host-galaxy redshifts,
as estimated from the spectroscopically confirmed non-Ia SNe in the \sns.

\begin{figure} [!t] 
\begin{center}
\includegraphics[width=5.75in]{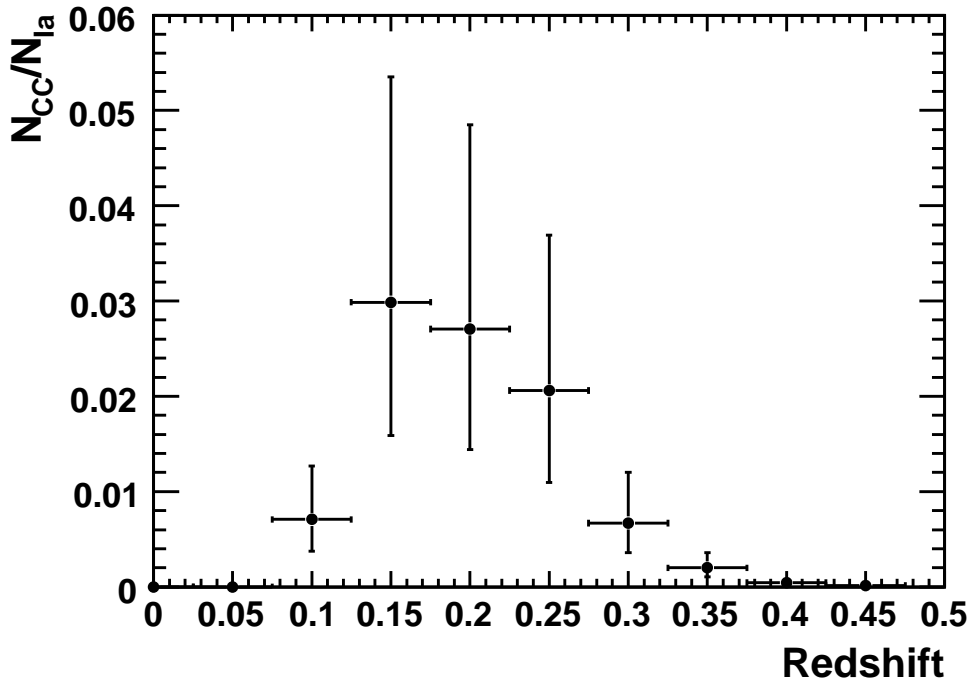}
\end{center}
\caption{Estimated core-collapse contamination fraction vs. redshift.
}
\label{fig:ccfracz}
\end{figure}

The estimated core-collapse contamination fraction as a function of redshift is 
shown in Figure~\ref{fig:ccfracz}, and the corresponding values
are given in Table~\ref{tab:midzratez1}.
The total estimated contamination of the SN sample
by non-Ia SNe to a redshift limit of 0.3 is
$2.04^{+1.61}_{-0.95}\%$. 
In the unbinned maximum-likelihood fits discussed in \S \ref{sec:ratevsz},
each photometric SN is given a weight according to the 
value of Eq.~\ref{eqn:ccfrac}.

\clearpage

\section{Survey Efficiency}
\label{sec:midzeffs}
The use of artificial SNe (fakes) in the survey discovery images 
and the SDSS-SN Monte Carlo (MC) simulation to 
determine the SN discovery and selection efficiency have been 
discussed in detail by \citet{Dilday_08a}.
For the MC simulation, for all observing epochs of the \sns, 
SN~Ia photometry is generated based 
on a SN Ia light-curve model (MLCS2k2 in the present analysis),
and the observing conditions corresponding to each epoch are
used to generate realistic photometry errors. 
Characteristics of the simulated SN sample, such as
distributions of time of maximum light, dust extinction, and 
intrinsic luminosity or decline rate, can be specified
in order to simulate a realistic SN sample and to 
investigate systematic effects of variations in the 
underlying distributions.

Here we discuss the effect on the SN discovery efficiency
of the modified selection procedure that uses statistics of the 
photometric-typing fits.
To study the SN discovery efficiency for this SN rate analysis, 
we used the SDSS-SN MC simulation to generate a sample of $\sim$17,000 
MC SNe Ia, which comprises $\sim$1000 SNe in each of 17 narrow
redshift bins in the range $0.025 < z < 0.4$. 
These MC SNe were filtered through a simulation of the 
search detection efficiency. That is, the efficiencies as a function
of signal-to-noise ratio determined from the fakes \citep{Dilday_08a}
were applied to the simulated
MC photometry. As in the search pipeline, a detection in at least
two of the $gri$ filters is required for the point to be considered
to have been detected and to be included in the fit.

The simulated search photometry was then fit with the same 
photometric-typing code used during the search, and the cuts outlined
in \S\ref{sec:midzsample} were applied. The resulting 
selection efficiencies are shown in Figure~\ref{fig:effz1}. 

As discussed by \citet{Sako_08}, the photometric typing can be done
with or without utilizing 
forced photometry (performing difference-imaging photometry at
known positions of transient objects) 
and with or without a prior on the SN redshift (from the host-galaxy
photometric or spectroscopic redshift measurement).
The selection cuts are applied to fits that
{\it do not} use forced photometry and {\it do not} use a prior
on the SN redshift. While it is evident from 
examining the photometric-typing fits during the 
SN search campaign that utilizing forced photometry
and/or a prior on the SN redshift in many cases 
improves the ability to distinguish SNe
Ia based on their search-photometry light curves, it is significantly
more difficult to model the selection function. 
The additional complications arise 
because forced photometry
was applied nonuniformly to the SDSS-SN candidates, and because modeling
the distribution of host-galaxy photometric redshifts and their errors is
nontrivial \citep{Oyaizu_08}. 

\subsection{Systematic Studies of the SN Discovery Efficiency}
\label{sec:midzeffsyst}

\begin{figure} [!t] 
\begin{center}
\plottwo{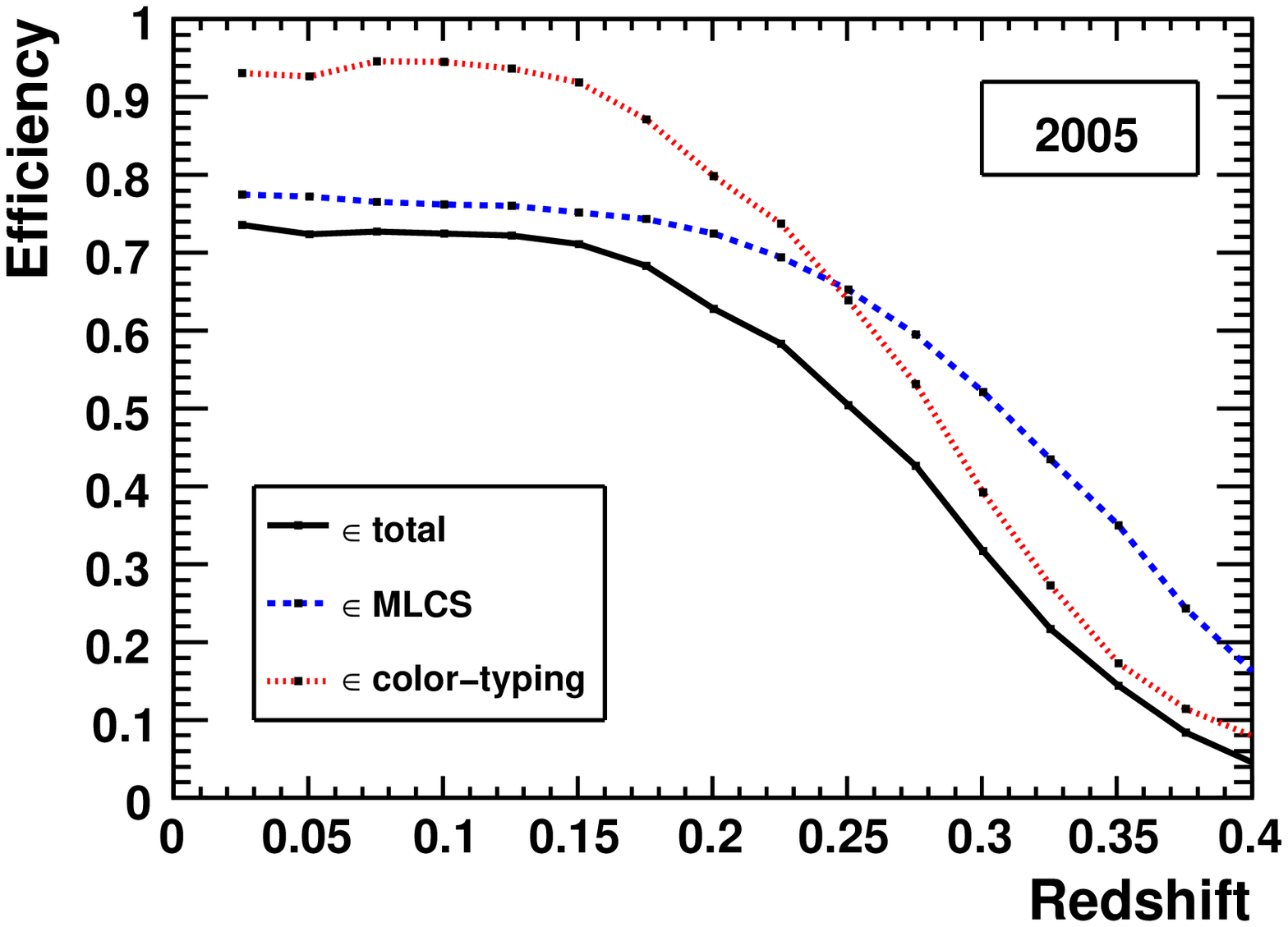}{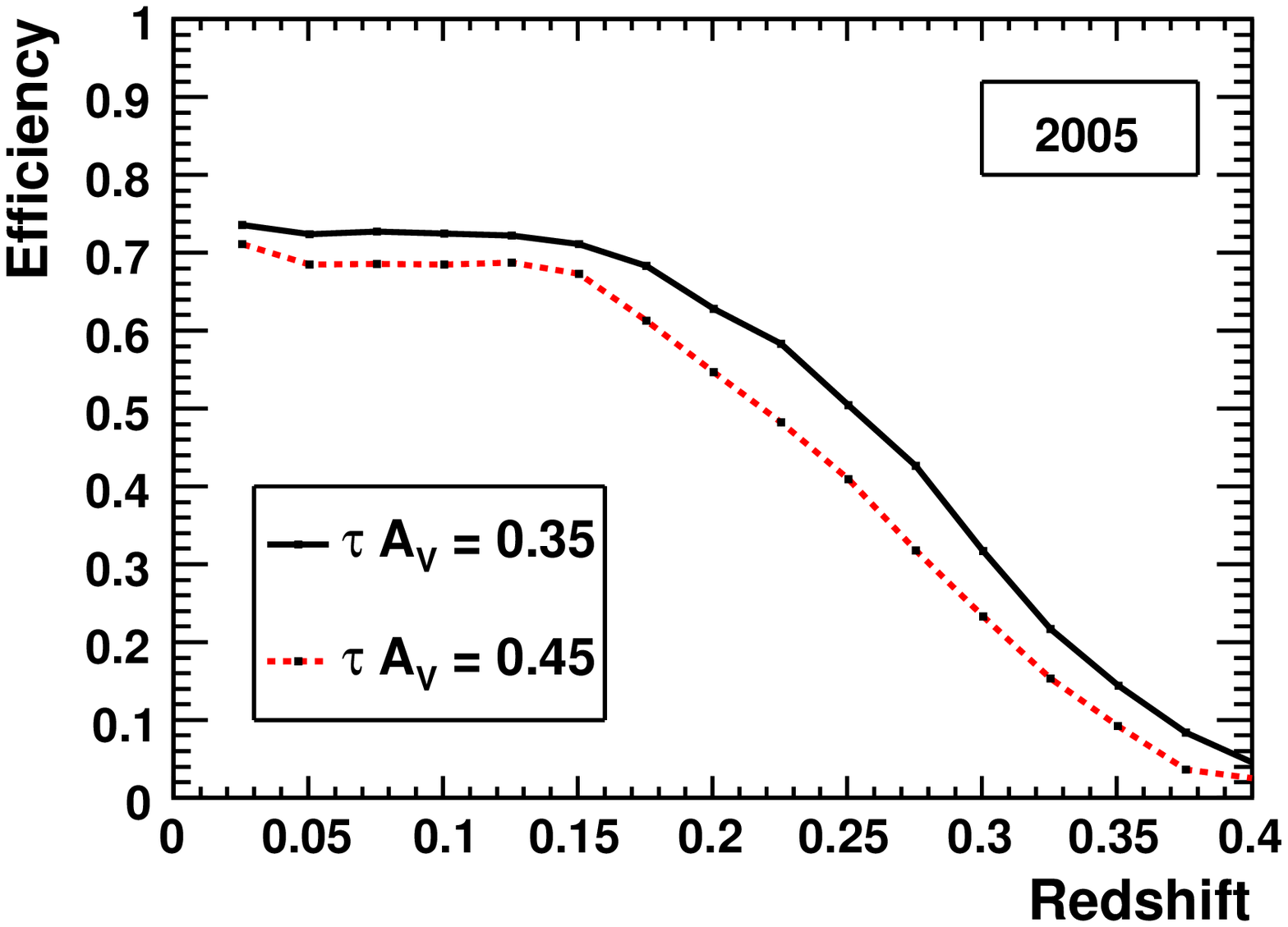}
\end{center}
\caption{
Results of studies of the SDSS-SN discovery efficiency, as a function of 
redshift, based on MC SN samples, for the 2005 observing season. 
The left panel shows the SN selection efficiency for the photometric-typing (red/dashed),
MLCS2k2 (blue/dashed), and combined (black/solid) selection criteria.
The right panel shows the efficiency for two different assumptions about the
distribution of the extinction parameter, $A_{V}$. The $A_V$ distribution
is assumed to have the form $e^{-A_{V}/\tau}$. The efficiency is shown
for $\tau = 0.35$ (default) and 
for $\tau = 0.45$ (1$\sigma$ upper limit). The values
for $\tau = 0.25$ (1$\sigma$ lower limit) are not shown since
they are indistinguishable from the $\tau = 0.35$ values. The efficiency 
curves for the 2006 and 2007 observing seasons show similar behavior.
 }
\label{fig:effz1}
\end{figure}

In \citet{Dilday_08a}, we considered the effect on the 
SN rate discovery efficiency of variation in the 
distribution of SN population parameters and found that
varying the distribution of extinction 
values had by far the largest effect. Here we 
repeat the systematic variation of the assumed extinction 
distribution with the modified selection procedure used in this paper.
We vary the input extinction distribution,
$p(A_{V}) \propto e^{-A_{V}/\tau}$,  with 
$\tau = 0.35 \pm 0.1$. 
The mean value and variation of $A_V$ are 
based upon investigation of the underlying $A_V$ distributions
presented in the SDSS-SN cosmology analysis \citep{Kessler_09}.
We find that for a low-extinction
set of SNe ($\tau = 0.25$), the efficiency 
differs negligibly from the default value of $\tau = 0.35$.
However, if the characteristic extinction is large ($\tau = 0.45$),
the efficiency differs markedly from the 
fiducial set of SNe, particularly for $z \gtrsim 0.2$.
Comparison of the efficiency between the fiducial and high-extinction
assumptions is shown in Figure \ref{fig:effz1} (right panel).
In Table \ref{tab:midzratez1} it can seen that the
systematic uncertainty in the SN rate due to 
uncertainty in the extinction distribution
becomes comparable to the statistical uncertainty 
for $z \approx 0.15$.

\section{SN Ia Rate Results}
\label{sec:results}

\subsection{Constant SN Ia Rate Model}
We first consider interpretations of the SN observations 
described above using a model of the SN rate that is 
constant as a function of redshift. 
In a constant-rate model, the volumetric rate is given by
\begin{equation}
r_V  = \frac{N}{\widetilde{V T \epsilon}},
\end{equation}

\noindent where
\begin{equation}
\label{eqn:midzeffthte}
\widetilde{V T \epsilon} = (\Theta T_{\earth} ) 
\int_{z_{min}}^{z_{max}}{\epsilon(z) ~\frac{d(VT/\Theta)}{dz}}\, dz,
\end{equation}

\noindent $N$ is the number of SNe in the sample, $T_{\earth}$ is the
observation time in the observer frame, $\Theta$ is the survey solid angle,
$\epsilon(z)$ is the SN discovery efficiency,
and $d(VT/\Theta)/dz$ 
is the element of volume multiplied by time per steradian
in the SN frame.
In the Friedmann-Robertson-Walker metric,
$d(VT/\Theta)/dz$ is given by

\begin{eqnarray}
d(VT/\Theta)/dz & = & u^2~\frac{du}{dz}~\frac{1}{1+z}, \\
u(z) & = & \int_{0}^{z}~\frac{c}{H(z')}\, dz'.
\end{eqnarray}

\begin{figure} [!t] 
\begin{center}
\includegraphics[width=5.75in]{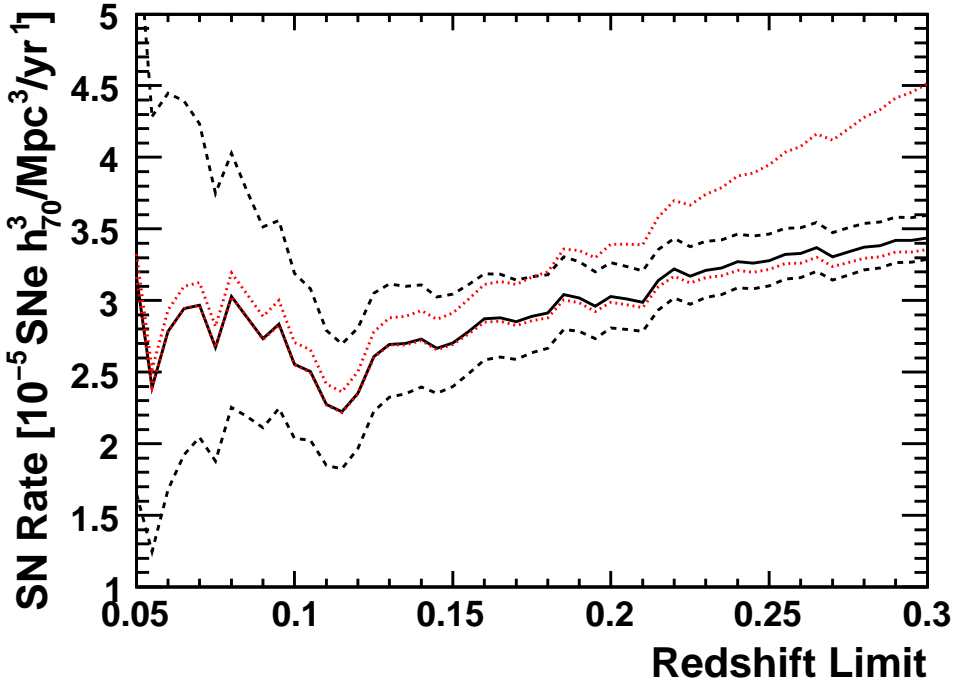}
\end{center}
\caption{
The SN Ia rate, assuming a constant-rate model, 
as a function of upper limit on the
redshift range (solid line). 
The black dashed lines denote the 1$\sigma$ statistical (Poisson) 
errors. The red dotted lines show estimates of the systematic errors,
which include uncertainty in the SN extinction distribution and 
estimation of contamination of the SN Ia sample from non-Ia SNe. 
}

\label{fig:ratez1}
\end{figure}

For the \sns, the Earth-frame observation time for the 
2005--2007 observing seasons are
89, 90, and 90 days, respectively. 
The solid angle covered is $\Theta = 0.08277 \times 0.98$ steradians.
As discussed by \citet{Dilday_08a}, the regions of the 
difference images that corresponded to the 
locations of bright stars and objects that showed 
variability in a previous year, and were thus unlikely to be SNe,
were excluded (masked) from the search for SNe. This masking accounts 
for the factor of 98\% in the computation of the
effective solid angle.
The value of the volumetric SN Ia rate, as a function of the upper redshift 
limit for the SN sample, and derived under the assumption of 
a constant-rate model, is shown in Figure \ref{fig:ratez1}. 
For example, if the upper redshift limit is chosen as $z=0.12$, as in 
\citet{Dilday_08a}, then the rate is determined to be
$r_V = (\lowzval^{+\lowzvalhistat+\lowzvalhisyst}_{-\lowzvallowstat-\lowzvallowsyst})$
$\times 10^{-5}$ 
$\rateunits$
(where $h_{70} = H_0/(70$~km~s$^{-1}$~Mpc$^{-1})$, and the quoted uncertainties
are statistical and systematic in that order),
based on 37 SNe Ia of which 36 are spectroscopically confirmed. 
This is lower than, although consistent with, the result found by
\citet{Dilday_08a},
$r_V = (\lowzvaloheight^{+\lowzvalhistatoheight+\lowzvalhisystoheight}_{-\lowzvallowstatoheight-\lowzvallowsystoheight})$
$\times 10^{-5}$ 
$\rateunits$. 
In relation to the mean low-redshift SN yield from the three years of the 
SDSS-II Supernova Survey, the first-year sample presented by 
\citet{Dilday_08a} represented an upward 
statistical fluctuation.
If the upper redshift limit is chosen as $z=0.15$, 
where the SN sample still has a high degree of completeness, 
then the SN Ia rate is determined to be 
$r_V = (\lowzvalonefive^{+\lowzvalonefivehistat+\lowzvalonefivehisyst}_{-\lowzvalonefivelowstat-\lowzvalonefivelowsyst})$
$\times 10^{-5}$
$\rateunits$, 
based on 79 SNe Ia of which 72 are spectroscopically confirmed.

\subsection{SN Ia Rate as a Function of Redshift}
\label{sec:ratevsz}
\begin{figure} [!t] 
\begin{center}
\includegraphics[width=5.75in]{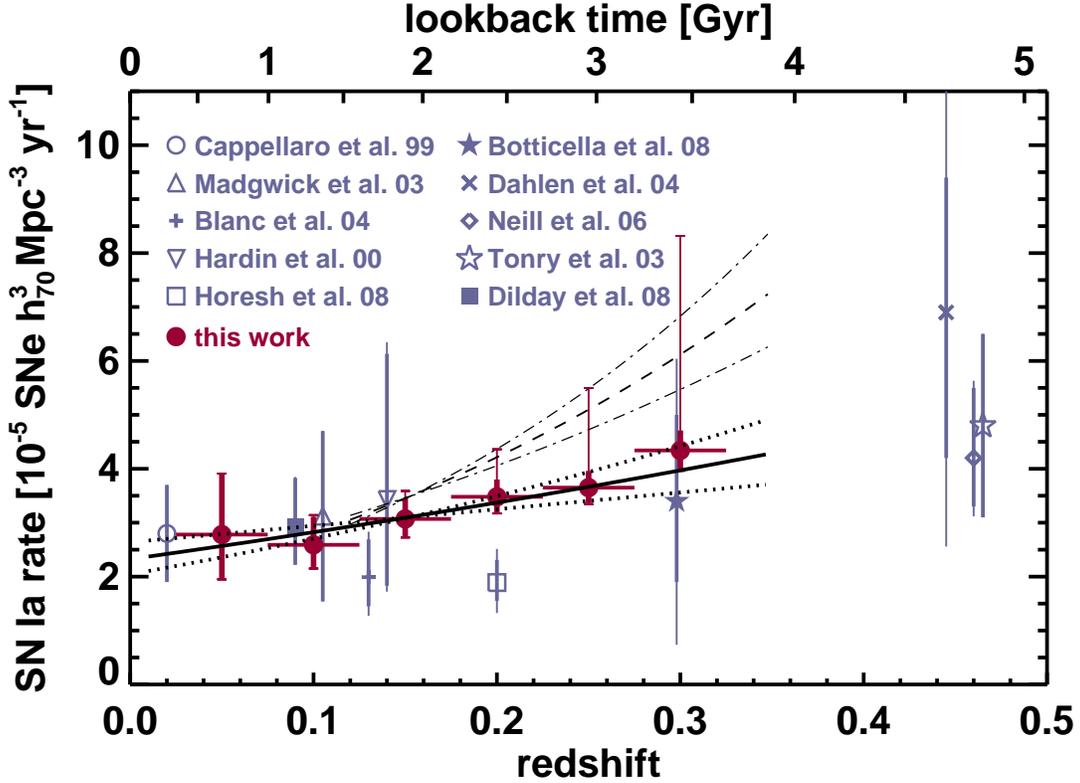}
\end{center}
\caption{
SN Ia rate as a function of redshift for the present 
work, along with a selection of measurements from the literature. 
For the measurements presented in this work, the 
redshift is the median redshift of 
running bins of size $\Delta z = 0.05$,
and the SN rate is computed assuming that the rate is 
constant in each bin. 
The thick error bars denote the statistical uncertainty while the 
thin error bars denote the systematic uncertainty. The solid line 
shows the best-fit power-law rate model, and the dotted lines the 
1$\sigma$ uncertainty of the best-fit model. 
The dashed line shows the best-fit power-law rate model (plotted
only for $z \ge 0.12$), 
assuming a larger mean value of dust extinction 
(\S \ref{sec:midzeffsyst}), and the dash-dotted line shows the
corresponding 1$\sigma$ uncertainty of the rate model.
Some of the SN Ia rate
measurements from the literature have been offset in redshift for 
clarity.
}
\label{fig:ratez2}
\end{figure}

In Figure \ref{fig:ratez2}, we plot the volumetric SN Ia rate in 
running bins of width 
$\Delta z = 0.05$. The rate in each bin is computed by assuming the rate
to be constant within the bin, which is 
a good approximation for the small bins considered. 
The SN rate values are listed in Table \ref{tab:midzratez1}.
It can be seen that the systematic uncertainty in the 
selection function (due to uncertainty in the extinction 
distribution) becomes much larger than the statistical uncertainty for 
$ z \gtrsim 0.2$. 

As in \citet{Dilday_08a}, we employ an unbinned maximum-likelihood method
to fit the SN rate data to models of the rate as a function of redshift.
To review the method, each SN redshift, $z^i$, is assumed to be drawn from a 
probability distribution,

\begin{equation}
p^i_z~dz = \langle N \rangle^{-1}~\Theta~T_{\earth}~\epsilon(z^i)~r_V(z^i)~d(VT/\Theta)/dz~dz,
\end{equation}
\noindent where $r_V(z)$ is the volumetric rate as a function of redshift, 
$\langle N \rangle$ is the mean number of expected SNe,
and the other symbols have been defined in Eq. \ref{eqn:midzeffthte}.
A likelihood function, $L$, can then be formed as a product of the individual
probabilities, $p_z^i$, multiplied by a Poisson distribution 
of the observed number of SNe, $N_{\mathrm{SNe}}$, 

\begin{equation}
L = \frac{\avgn^{\nobs}~e^{-\avgn}}{(\nobs)!}~\Pi_{i=1}^{i=\nobs}~p_z^i.
\end{equation}

\noindent When the SN redshifts are determined photometrically, the 
uncertainties in the measurements are significant, and the individual 
probabilities must be 
modified as

\begin{equation}
p^i_z = \langle N \rangle^{-1}~\Theta~T_{\earth}~\int_{-\infty}^{\infty}
~\epsilon(z')~r_V(z')~d(VT/\Theta)/dz'~\rho(z'|z^i) \, dz',
\end{equation}

\noindent where $\rho(z'|z^i)$ is the probability that a SN at redshift $z^i$
will have a measured redshift $z'$. 
We assume a Gaussian form for $\rho$,

\begin{equation}
\rho(z'|z^i) = \frac{1}{\sqrt{2\pi}\sigma_z}~e^{-(z'-z^i)^2/2\sigma_z^2},
\end{equation}

\noindent where $\sigma_z$ is the uncertainty in the SN redshift,
as determined by the SN photo-$z$ fits.
The parameters of the SN rate model, $r_V$, are then estimated by minimizing the
negative log of the likelihood function. This procedure automatically accounts for the
bias in the observed SN redshift distribution described in \S \ref{sec:dndzbias}.
As discussed in \S \ref{sec:contamination}, contamination of the 
photometric SN sample by non-Ia SNe is accounted for by weighting each
photometric SN according to Eq.~\ref{eqn:ccfrac}.

Using the maximum-likelihood formalism described above, we consider an 
empirical power-law model of the SN rate as a function of redshift,
$r_V(z) = A_p \, ((1+z)/(1+z_{0}))^\nu$. 
The reference redshift is $z_{0} = 0.21$, and the SNe Ia used in the fit
are those with $z \le 0.3$.
Our minimization and error analysis are 
performed with the MINUIT software package, 
and using the MINOS procedure for asymmetric error estimation \citep{minuit}.
For the power-law rate model, the maximum-likelihood estimates of the 
model parameters are 
$A_p = (\apval^{+\apvalhi}_{-\apvallo}) \times 10^{-5}$ 
$\rateunits$
and 
$\nu = \nuval^{+\nuvalhi}_{-\nuvallo}$, with 
correlation coefficient $\rho = \powcorr$.
The uncertainties quoted 
above are the 1$\sigma$ statistical errors, defined as the 
change in the parameter values such that the log-likelihood function
changes by 0.5.
As discussed in \S \ref{sec:midzeffsyst}, uncertainty in the 
extinction distribution for SNe Ia has a significant impact on the
survey efficiency, as inferred from MC simulations, particularly
for $z \gtrsim 0.2$ (Fig.~\ref{fig:ratez2}).
Assuming a larger mean value 
($\langle A_V \rangle = 0.45$ mag)
for the extinction parameter, $A_V$, 
and reevaluating the maximum-likelihood estimates for the 
rate model parameters, results in 
$A_p = (\apvalav^{+\apvalhiav}_{-\apvalloav}) \times 10^{-5}$ 
$\rateunits$ and 
$\nu = \nuvalav^{+\nuvalhiav}_{-\nuvalloav}$, 
for the parameters of the 
power-law rate model.

The power-law SN rate models, along with a selection of 
SN Ia rate measurements from the literature, 
are shown in Figure \ref{fig:ratez2}. 
SN Ia rate measurements from the SDSS-II Supernova Survey, computed
in bins of width $\Delta z = 0.05$ (Table \ref{tab:midzratez1}),
are also shown. We emphasize that the binned SDSS-II Supernova Survey points 
are shown for convenience and ease of comparison, and that
the power-law rate models are derived from the unbinned 
maximum-likelihood method (\S \ref{sec:ratevsz}).
This plot illustrates the much greater statistical precision
of the \sns~in comparison to previous SN Ia rate measurements
in the redshift range $ 0.1 \lesssim z \lesssim 0.3$. 
Additionally, the \sns~SN Ia rate measurements in the redshift range 
$0.1 \lesssim z \lesssim 0.2$ have a relatively small
systematic uncertainty, and constrain the SN Ia rate to 
$\sim$10--20\% {\it total} uncertainty.
For $z \gtrsim 0.2$, the systematic uncertainty in the
\sns~SN Ia rate measurements becomes much greater than the statistical
uncertainty.
Despite this large systematic uncertainty,
the \sns~SN Ia rate measurements 
have precision comparable to that of the best existing measurements.
Finally, we note that the direction of the systematic uncertainty 
in the \sns~SN Ia rate measurements is toward an increase in the SN rate.
Therefore, the SN Ia rate measurements presented here provide quite robust 
lower limits on the SN Ia rate at $z \lesssim 0.3$.

\section{Conclusions}
\label{sec:midzconclude}

We have measured the volumetric SN Ia rate based on the \sns~using
a much larger sample of SNe Ia and a higher redshift limit 
than was discussed by \citet{Dilday_08a}. 
The sample of SNe considered comprises 
$\nsnezltohthree$ 
SNe Ia at $z \lesssim 0.3$.
The low-redshift portion of the SN Ia sample has a high 
degree of spectroscopic completeness, while the
large redshift range covered enables measurement of the 
redshift dependence of the SN Ia rate based on the 
\sns~data alone. Fitting a power-law model to the SN Ia rate,
$r_V(z) = A_p \, ((1+z)/(1+z_{0}))^\nu$, and assuming 
a distribution for dust extinction as in \citet{Kessler_09}, 
we find $\nu = \nuval^{+\nuvalhi}_{-\nuvallo}$.
Assuming a larger mean value of dust extinction
we find $\nu = \nuvalav^{+\nuvalhiav}_{-\nuvalloav}$.

The results presented here represent an order of magnitude 
improvement in the statistics for SN Ia rate measurements in the
same redshift range and solidify the SN Ia rate constraints 
for $z \lesssim 0.3$. When combined with improved measurements of the 
cosmic star formation history, the SN Ia rate measurements presented here
can be used to place improved constraints on SN Ia progenitor models.

\clearpage
\begin{deluxetable}{rcrrcc}
\tablecolumns{6}
\tabletypesize{\small}
\singlespace
\tablewidth{0pc} 
\tablecaption{Confirmed SNe Ia in the Rate Sample
\label{tab:snerates2_conf}
}
\tablehead{
\colhead{SN ID} & 
\colhead{IAU Name} &    
\colhead{$\alpha_{\mathrm{J2000}}$} &    
\colhead{$\delta_{\mathrm{J2000}}$} & 
\colhead{Redshift} &
\colhead{Fitprob} \\ 

\colhead{} & 
\colhead{} & 
\colhead{[degrees]} &    
\colhead{[degrees]} & 
\colhead{} &
\colhead{}

}
\renewcommand{\arraystretch}{1.5}
\startdata
762 & 2005eg & 15.53518 & -0.87907 & 0.191 & 0.945 \\
1032 & 2005ez & 46.79565 & +1.11952 & 0.130 & 0.541 \\
1112 & 2005fg & 339.01761 & -0.37527 & 0.258 & 0.792 \\
1166 & \nodata & 9.35560 & +0.97320 & 0.382 & 0.992 \\
1241 & 2005ff & 337.67249 & -0.77664 & 0.090 & 0.988 \\
1253 & 2005fd & 323.79895 & +0.16305 & 0.262 & 0.834 \\
1371 & 2005fh & 349.37375 & +0.42929 & 0.119 & 0.995 \\
1580 & 2005fb & 45.32296 & -0.64412 & 0.183 & 1.000 \\
1688 & \nodata & 321.35767 & +0.32447 & 0.359 & 0.183 \\
2017 & 2005fo & 328.94327 & +0.59343 & 0.262 & 0.972 \\
2031 & 2005fm & 312.04312 & -1.17149 & 0.153 & 0.894 \\
2165 & 2005fr & 17.09165 & -0.09639 & 0.288 & 0.930 \\
2246 & 2005fy & 50.09031 & -0.88564 & 0.195 & 0.904 \\
2308 & 2005ey & 34.27273 & +0.28020 & 0.148 & 0.960 \\
2330 & 2005fp & 6.80699 & +1.12053 & 0.213 & 0.841 \\
2372 & 2005ft & 40.52065 & -0.54088 & 0.181 & 0.996 \\
2422 & 2005fi & 1.99444 & +0.63811 & 0.265 & 0.867 \\
2440 & 2005fu & 42.63359 & +0.80774 & 0.193 & 0.859 \\
2533 & 2005fs & 31.22063 & -0.32655 & 0.340 & 0.995 \\
2561 & 2005fv & 46.34335 & +0.85829 & 0.118 & 1.000 \\
2635 & 2005fw & 52.70425 & -1.23815 & 0.144 & 0.994 \\
2689 & 2005fa & 24.90027 & -0.75879 & 0.162 & 0.230 \\
2789 & 2005fx & 344.20142 & +0.40102 & 0.290 & 0.231 \\
2916 & 2005fz & 315.92169 & +0.56948 & 0.124 & 0.341 \\
2943 & 2005go & 17.70491 & +1.00784 & 0.265 & 0.744 \\
2992 & 2005gp & 55.49692 & -0.78269 & 0.127 & 0.385 \\
3080 & 2005ga & 16.93231 & -1.03955 & 0.175 & 0.999 \\
3087 & 2005gc & 20.40666 & -0.97730 & 0.166 & 1.000 \\
3199 & 2005gs & 333.29257 & +1.05048 & 0.251 & 0.817 \\
3241 & 2005gh & 312.65143 & -0.35416 & 0.259 & 0.219 \\
3331 & 2005ge & 34.56134 & +0.79646 & 0.213 & 0.993 \\
3377 & 2005gr & 54.15606 & +1.07909 & 0.245 & 0.999 \\
3451 & 2005gf & 334.06921 & +0.70805 & 0.250 & 0.814 \\
3452 & 2005gg & 334.67139 & +0.63906 & 0.230 & 0.957 \\
3592 & 2005gb & 19.05240 & +0.79183 & 0.087 & 1.000 \\
3901 & 2005ho & 14.85039 & +0.00252 & 0.063 & 1.000 \\
4046 & 2005gw & 354.49823 & +0.64201 & 0.277 & 0.955 \\
4241 & 2005gu & 12.23758 & -0.90583 & 0.332 & 0.176 \\
4577 & 2005gv & 38.47543 & +0.28062 & 0.363 & 0.994 \\
4679 & 2005gy & 21.52825 & +0.67678 & 0.332 & 0.413 \\
5103 & 2005gx & 359.88437 & +0.73711 & 0.162 & 0.993 \\
5350 & 2005hp & 307.21909 & -0.77928 & 0.175 & 0.999 \\
5395 & 2005hr & 49.64084 & +0.12327 & 0.117 & 1.000 \\
5533 & 2005hu & 328.66986 & +0.41335 & 0.220 & 0.733 \\
5549 & 2005hx & 3.25048 & +0.24814 & 0.121 & 0.997 \\
5550 & 2005hy & 3.59819 & +0.33297 & 0.156 & 0.988 \\
5635 & 2005hv & 333.18271 & -0.03503 & 0.179 & 0.943 \\
5717 & 2005ia & 17.89588 & -0.00597 & 0.252 & 0.999 \\
5736 & 2005jz & 22.86264 & -0.63166 & 0.253 & 0.962 \\
5737 & 2005ib & 22.85692 & -0.60344 & 0.393 & 0.890 \\
5751 & 2005hz & 11.63408 & +0.83812 & 0.131 & 0.999 \\
5844 & 2005ic & 327.78616 & -0.84302 & 0.311 & 0.837 \\
5916 & 2005is & 5.43728 & -0.32510 & 0.175 & 0.923 \\
5957 & 2005ie & 34.76051 & -0.27292 & 0.280 & 0.840 \\
5994 & 2005ht & 312.60245 & -0.16795 & 0.187 & 0.993 \\
6057 & 2005if & 52.55355 & -0.97467 & 0.067 & 1.000 \\
6108 & 2005ih & 1.80657 & +0.34890 & 0.259 & 0.993 \\
6127 & 2005iw & 337.32391 & -0.09245 & 0.280 & 0.696 \\
6137 & 2005iv & 307.93607 & +0.24473 & 0.300 & 0.964 \\
6192 & 2005jy & 348.46494 & +1.25689 & 0.272 & 0.837 \\
6196 & 2005ig & 337.63107 & -0.50279 & 0.281 & 0.290 \\
6249 & 2005ii & 3.26547 & -0.62019 & 0.294 & 0.897 \\
6295 & 2005js & 23.67295 & -0.60544 & 0.080 & 0.005 \\
6304 & 2005jk & 26.49760 & +1.19589 & 0.191 & 0.896 \\
6315 & 2005ix & 310.48273 & +1.09191 & 0.267 & 0.828 \\
6406 & 2005ij & 46.08860 & -1.06301 & 0.125 & 1.000 \\
6422 & 2005id & 349.13870 & -0.66326 & 0.184 & 0.954 \\
6558 & 2005hj & 21.70165 & -1.23814 & 0.057 & 0.998 \\
6649 & 2005jd & 34.27582 & +0.53471 & 0.314 & 0.949 \\
6699 & 2005ik & 322.81491 & -1.05707 & 0.311 & 0.894 \\
6773 & 2005iu & 305.06503 & +0.21731 & 0.090 & 0.009 \\
6780 & 2005iz & 328.06851 & +0.26698 & 0.202 & 0.988 \\
6933 & 2005jc & 11.35168 & +1.07547 & 0.213 & 0.951 \\
6936 & 2005jl & 323.23380 & -0.69986 & 0.181 & 0.772 \\
7143 & 2005jg & 345.26233 & -0.20746 & 0.304 & 0.981 \\
7147 & 2005jh & 350.01831 & -0.05557 & 0.111 & 0.080 \\
7243 & 2005jm & 328.07904 & +0.47186 & 0.204 & 0.945 \\
7335 & 2005kn & 318.88513 & -0.35543 & 0.198 & 0.850 \\
7473 & 2005ji & 4.32628 & -0.25736 & 0.216 & 0.942 \\
7475 & 2005jn & 4.75338 & -0.28156 & 0.322 & 0.911 \\
7512 & 2005jo & 52.09022 & -0.32617 & 0.219 & 0.708 \\
7779 & 2005jw & 310.08017 & -0.00723 & 0.381 & 0.406 \\
7847 & 2005jp & 32.45985 & -0.06185 & 0.212 & 0.838 \\
7876 & 2005ir & 19.18237 & +0.79452 & 0.076 & 0.298 \\
7947 & 2005jj & 314.18600 & +0.40829 & 0.368 & 0.563 \\
8030 & 2005jv & 40.20864 & +0.99305 & 0.422 & 0.997 \\
8046 & 2005ju & 39.11671 & +0.51118 & 0.259 & 0.998 \\
8213 & 2005ko & 357.52081 & -0.92148 & 0.185 & 0.839 \\
8495 & 2005mi & 335.26102 & -0.74820 & 0.214 & 0.268 \\
8598 & 2005jt & 42.66724 & -0.06600 & 0.361 & 0.991 \\
8707 & 2005mh & 41.23612 & +0.20358 & 0.395 & 0.877 \\
8719 & 2005kp & 7.72141 & -0.71893 & 0.118 & 0.298 \\
9032 & 2005le & 337.88452 & -0.49365 & 0.254 & 0.991 \\
9045 & 2005kq & 347.83704 & -0.60869 & 0.389 & 0.788 \\
9467 & 2005lh & 328.95132 & +1.18071 & 0.218 & 0.413 \\
10550 & 2005lf & 349.67532 & -1.20496 & 0.300 & 0.916 \\
12780 & 2006eq & 322.15454 & +1.22804 & 0.049 & 0.997 \\
12853 & 2006ey & 316.76547 & +0.72302 & 0.169 & 0.688 \\
12855 & 2006fk & 330.25549 & +0.71614 & 0.172 & 0.006 \\
12856 & 2006fl & 332.86542 & +0.75589 & 0.172 & 0.990 \\
12860 & 2006fc & 323.69412 & +1.17579 & 0.122 & 0.999 \\
12874 & 2006fb & 353.96448 & -0.17724 & 0.245 & 0.795 \\
12898 & 2006fw & 26.79302 & -0.14708 & 0.083 & 0.997 \\
12930 & 2006ex & 309.68271 & -0.47460 & 0.147 & 0.999 \\
12950 & 2006fy & 351.66742 & -0.84032 & 0.083 & 0.987 \\
12972 & 2006ft & 7.95851 & -0.38302 & 0.261 & 0.993 \\
12977 & 2006gh & 13.69553 & -0.25095 & 0.248 & 1.000 \\
13025 & 2006fx & 341.56720 & +0.41579 & 0.224 & 0.925 \\
13038 & 2006gn & 347.82672 & +0.50451 & 0.104 & 0.997 \\
13044 & 2006fm & 332.54269 & +0.50323 & 0.126 & 1.000 \\
13070 & 2006fu & 357.78500 & -0.74645 & 0.199 & 0.820 \\
13072 & 2006fi & 334.95929 & +0.02439 & 0.231 & 0.817 \\
13135 & 2006fz & 4.17214 & -0.42463 & 0.105 & 0.756 \\
13136 & 2006go & 6.14062 & -0.27912 & 0.372 & 0.174 \\
13152 & 2006gg & 7.05201 & +0.11793 & 0.203 & 0.935 \\
13174 & 2006ga & 13.23463 & +0.44774 & 0.236 & 0.570 \\
13305 & 2006he & 331.10043 & +0.69111 & 0.214 & 0.392 \\
13354 & 2006hr & 27.56474 & -0.88735 & 0.158 & 0.681 \\
13411 & \nodata & 315.18991 & +0.19162 & 0.163 & 0.967 \\
13506 & 2006hg & 25.24323 & -0.72791 & 0.245 & 0.013 \\
13511 & 2006hh & 40.61224 & -0.79419 & 0.238 & 0.176 \\
13578 & 2006hc & 17.39463 & +0.70404 & 0.229 & 0.713 \\
13641 & 2006hf & 345.21866 & -0.98122 & 0.219 & 0.393 \\
13689 & \nodata & 4.01602 & +0.80747 & 0.252 & 0.826 \\
13727 & 2006hj & 317.58804 & +0.93256 & 0.226 & 0.451 \\
13736 & 2006hv & 336.83325 & +1.03089 & 0.150 & 0.999 \\
13757 & 2006hk & 350.12305 & -1.15792 & 0.289 & 0.641 \\
13796 & 2006hl & 350.69150 & +0.53280 & 0.148 & 1.000 \\
13835 & 2006hp & 6.05959 & -0.24832 & 0.248 & 0.915 \\
13894 & 2006jh & 1.69050 & -0.03686 & 0.125 & 0.999 \\
13934 & 2006jg & 342.11084 & -0.43543 & 0.330 & 0.437 \\
13956 & 2006hi & 20.94173 & +0.81663 & 0.262 & 0.020 \\
14019 & 2006ki & 316.64233 & -0.64799 & 0.216 & 0.563 \\
14024 & 2006ht & 318.19852 & +0.91627 & 0.146 & 0.707 \\
14108 & 2006hu & 53.59454 & -1.12330 & 0.133 & 0.827 \\
14157 & 2006kj & 51.13660 & +1.02230 & 0.212 & 0.023 \\
14212 & 2006iy & 330.47003 & +1.04491 & 0.205 & 0.745 \\
14261 & 2006jk & 328.24030 & +0.25372 & 0.286 & 0.963 \\
14279 & 2006hx & 18.48869 & +0.37156 & 0.045 & 0.395 \\
14284 & 2006ib & 49.04921 & -0.60105 & 0.181 & 0.237 \\
14298 & 2006jj & 314.89502 & +1.22317 & 0.270 & 0.935 \\
14331 & 2006kl & 7.88862 & -0.13590 & 0.221 & 0.619 \\
14377 & 2006hw & 48.26427 & -0.47171 & 0.139 & 0.744 \\
14421 & 2006ia & 31.82975 & +1.25197 & 0.175 & 0.005 \\
14437 & 2006hy & 332.08087 & -1.19642 & 0.149 & 0.803 \\
14481 & 2006lj & 2.68159 & +0.20113 & 0.244 & 0.482 \\
14735 & 2006km & 35.15793 & +0.34831 & 0.301 & 0.858 \\
14782 & 2006jp & 314.23422 & -0.27913 & 0.160 & 0.048 \\
14816 & 2006ja & 336.71625 & +0.50600 & 0.107 & 0.020 \\
14846 & 2006jn & 7.66319 & +0.14148 & 0.225 & 0.991 \\
14871 & 2006jq & 54.27705 & +0.00928 & 0.128 & 0.997 \\
14979 & 2006jr & 54.94641 & +0.99273 & 0.177 & 0.981 \\
14984 & 2006js & 313.83350 & -0.09283 & 0.197 & 0.619 \\
15002 & 2006ko & 22.24966 & +0.76984 & 0.356 & 0.725 \\
15057 & 2006md & 17.88118 & +0.40941 & 0.299 & 0.105 \\
15129 & 2006kq & 318.90228 & -0.32151 & 0.199 & 0.025 \\
15132 & 2006jt & 329.70016 & +0.19764 & 0.144 & 0.853 \\
15136 & 2006ju & 351.16245 & -0.71842 & 0.149 & 0.023 \\
15161 & 2006jw & 35.84275 & +0.81893 & 0.250 & 0.212 \\
15171 & 2006kb & 304.79251 & -1.06454 & 0.134 & 0.995 \\
15201 & 2006ks & 337.51941 & +0.00359 & 0.208 & 0.352 \\
15203 & 2006jy & 15.73466 & +0.18302 & 0.204 & 0.989 \\
15213 & 2006lk & 53.01909 & -0.10022 & 0.311 & 0.982 \\
15219 & 2006ka & 34.61100 & +0.22658 & 0.248 & 0.879 \\
15222 & 2006jz & 2.85320 & +0.70267 & 0.199 & 0.995 \\
15229 & 2006kr & 4.83195 & +1.09059 & 0.227 & 0.844 \\
15234 & 2006kd & 16.95826 & +0.82809 & 0.136 & 0.978 \\
15254 & 2006oy & 313.49286 & -0.36011 & 0.201 & 0.186 \\
15259 & 2006kc & 337.54407 & -0.40788 & 0.210 & 0.725 \\
15287 & 2006kt & 323.95966 & -1.05750 & 0.274 & 0.909 \\
15301 & 2006lo & 323.57986 & +0.58907 & 0.296 & 0.738 \\
15354 & 2006lp & 6.77357 & -0.12606 & 0.222 & 0.199 \\
15356 & 2006lm & 335.05325 & +0.40984 & 0.275 & 0.162 \\
15365 & 2006ku & 354.55658 & +1.24900 & 0.188 & 0.963 \\
15369 & 2006ln & 348.83286 & -0.56267 & 0.232 & 0.881 \\
15421 & 2006kw & 33.74157 & +0.60240 & 0.185 & 0.981 \\
15425 & 2006kx & 55.56103 & +0.47820 & 0.160 & 0.640 \\
15433 & 2006mt & 14.87962 & -0.25662 & 0.221 & 0.879 \\
15440 & 2006lr & 39.72062 & +0.09002 & 0.262 & 0.801 \\
15443 & 2006lb & 49.86736 & -0.31813 & 0.182 & 0.617 \\
15453 & 2006ky & 319.66821 & -1.02438 & 0.184 & 0.999 \\
15456 & 2006ll & 331.86716 & -0.90348 & 0.382 & 0.626 \\
15459 & 2006la & 340.70139 & -0.90184 & 0.127 & 0.832 \\
15461 & 2006kz & 326.84741 & -0.49477 & 0.180 & 0.931 \\
15466 & 2006mz & 317.64496 & -0.12329 & 0.246 & 0.930 \\
15467 & \nodata & 320.01984 & -0.17753 & 0.210 & 0.984 \\
15508 & 2006ls & 27.16886 & -0.57666 & 0.147 & 0.997 \\
15583 & 2006mv & 37.73095 & +0.94621 & 0.175 & 0.911 \\
15584 & 2006nt & 43.49536 & +0.98691 & 0.282 & 0.994 \\
15648 & 2006ni & 313.71829 & -0.19488 & 0.175 & 0.316 \\
15674 & 2006nu & 340.82910 & +0.26279 & 0.197 & 0.796 \\
15704 & 2006nh & 40.21070 & +0.65870 & 0.365 & 0.906 \\
15776 & 2006na & 32.82938 & -0.99828 & 0.305 & 0.046 \\
15868 & 2006pa & 38.09978 & -0.71369 & 0.242 & 0.923 \\
15872 & 2006nb & 36.72244 & -0.32791 & 0.185 & 0.941 \\
15897 & 2006pb & 11.68145 & -1.03294 & 0.175 & 0.051 \\
15901 & 2006od & 31.97622 & -0.53548 & 0.205 & 0.413 \\
16000 & 2006nj & 21.11753 & +0.07430 & 0.390 & 0.964 \\
16021 & 2006nc & 13.84359 & -0.38889 & 0.124 & 0.457 \\
16032 & 2006nk & 44.06910 & -0.41082 & 0.204 & 0.735 \\
16072 & 2006nv & 3.12437 & -0.97737 & 0.287 & 0.896 \\
16073 & 2006of & 8.10770 & -1.05408 & 0.153 & 0.955 \\
16093 & 2006oe & 350.36243 & +1.13236 & 0.335 & 0.994 \\
16100 & 2006nl & 30.43619 & -1.03248 & 0.195 & 0.559 \\
16185 & 2006ok & 16.86800 & -0.26945 & 0.097 & 0.987 \\
16211 & 2006nm & 348.16397 & +0.26679 & 0.311 & 0.928 \\
17168 & 2007ik & 339.72379 & -1.16716 & 0.184 & 0.995 \\
17186 & 2007hx & 31.61272 & -0.89963 & 0.080 & 0.989 \\
17332 & 2007jk & 43.77335 & -0.14750 & 0.183 & 0.434 \\
17340 & 2007kl & 41.21201 & +0.36474 & 0.257 & 0.945 \\
17366 & 2007hz & 315.78723 & -1.02926 & 0.139 & 0.993 \\
17389 & 2007ih & 323.29477 & -0.96028 & 0.171 & 0.674 \\
17435 & 2007ka & 20.34463 & -0.01488 & 0.218 & 0.310 \\
17497 & 2007jt & 37.13649 & -1.04221 & 0.145 & 0.991 \\
17552 & 2007jl & 322.32047 & -1.00328 & 0.254 & 0.847 \\
17568 & 2007kb & 313.10287 & +0.27759 & 0.145 & 0.999 \\
17629 & 2007jw & 30.63637 & -1.08924 & 0.137 & 0.324 \\
17745 & 2007ju & 2.96016 & -0.33928 & 0.064 & 1.000 \\
17784 & 2007jg & 52.46166 & +0.05676 & 0.037 & 1.000 \\
17791 & 2007kp & 332.37314 & +0.73781 & 0.286 & 0.526 \\
17801 & 2007ko & 316.09314 & -0.89806 & 0.206 & 0.572 \\
17809 & 2007kr & 6.36447 & -0.83962 & 0.282 & 0.934 \\
17811 & 2007ix & 12.87864 & -0.94747 & 0.213 & 0.877 \\
17825 & 2007je & 32.94703 & -0.91251 & 0.161 & 1.000 \\
17875 & 2007jz & 20.98331 & +1.25488 & 0.232 & 0.423 \\
17880 & 2007jd & 44.97227 & +1.16063 & 0.073 & 0.934 \\
17884 & 2007kt & 27.59980 & +1.17200 & 0.239 & 0.717 \\
17886 & 2007jh & 54.00627 & +1.10330 & 0.041 & 0.940 \\
18030 & 2007kq & 4.93285 & -0.40022 & 0.156 & 0.170 \\
18091 & 2007ku & 23.36750 & +0.52460 & 0.372 & 0.300 \\
18241 & 2007ks & 312.38766 & -0.76182 & 0.095 & 0.992 \\
18298 & 2007li & 18.26664 & -0.54014 & 0.120 & 0.416 \\
18323 & 2007kx & 3.42863 & +0.65206 & 0.155 & 0.893 \\
18325 & 2007mv & 8.90570 & +0.36997 & 0.255 & 0.372 \\
18375 & 2007lg & 11.51640 & -0.01046 & 0.110 & 0.784 \\
18415 & 2007la & 337.47827 & +1.05851 & 0.131 & 0.993 \\
18451 & 2007mt & 26.48567 & -0.21765 & 0.408 & 0.711 \\
18456 & 2007lk & 29.45877 & -0.39821 & 0.219 & 0.102 \\
18463 & 2007kv & 17.56589 & +0.47195 & 0.268 & 0.997 \\
18466 & 2007lm & 48.41846 & +0.62995 & 0.213 & 0.988 \\
18485 & 2007nu & 47.95904 & -0.69255 & 0.282 & 0.884 \\
18486 & 2007ln & 55.17992 & +1.00300 & 0.240 & 0.994 \\
18602 & 2007lo & 338.98349 & +0.60916 & 0.138 & 0.655 \\
18604 & 2007lp & 340.92081 & +0.42131 & 0.176 & 0.925 \\
18612 & 2007lc & 12.28788 & +0.59688 & 0.115 & 0.796 \\
18650 & 2007lt & 328.44720 & +0.01500 & 0.113 & 1.000 \\
18721 & 2007mu & 3.07724 & -0.07738 & 0.403 & 0.948 \\
18740 & 2007mc & 16.85519 & +1.04368 & 0.157 & 0.995 \\
18749 & 2007mb & 12.54733 & +0.67532 & 0.189 & 0.754 \\
18768 & 2007lh & 17.21666 & +1.19777 & 0.198 & 0.550 \\
18787 & 2007mf & 29.72962 & -1.02722 & 0.207 & 0.998 \\
18804 & 2007me & 25.26560 & -0.44851 & 0.205 & 0.941 \\
18807 & 2007mg & 46.64086 & +0.79320 & 0.158 & 1.000 \\
18835 & 2007mj & 53.68503 & +0.35546 & 0.123 & 1.000 \\
18855 & 2007mh & 48.63231 & +0.26975 & 0.128 & 1.000 \\
18890 & 2007mm & 16.44440 & -0.75890 & 0.066 & 0.970 \\
18903 & 2007lr & 12.25138 & -0.32410 & 0.156 & 0.991 \\
18909 & 2007lq & 5.78274 & +0.98334 & 0.228 & 0.816 \\
18927 & 2007nt & 46.68240 & -0.75412 & 0.213 & 0.696 \\
18940 & 2007sb & 10.34886 & +0.41188 & 0.212 & 0.424 \\
18945 & 2007nd & 10.07834 & -1.03742 & 0.263 & 0.839 \\
18965 & 2007ne & 13.50917 & +1.06891 & 0.207 & 0.754 \\
19002 & 2007nh & 42.61525 & -0.55126 & 0.263 & 0.999 \\
19008 & 2007mz & 331.96326 & -1.07002 & 0.232 & 0.940 \\
19027 & 2007my & 328.88412 & -0.37204 & 0.293 & 0.325 \\
19051 & 2007nb & 351.37521 & +0.42302 & 0.277 & 0.053 \\
19067 & 2007oq & 325.62805 & +0.98454 & 0.339 & 0.494 \\
19101 & 2007ml & 7.97267 & +0.13823 & 0.187 & 0.836 \\
19128 & 2007lw & 354.20370 & -0.78234 & 0.287 & 0.930 \\
19149 & 2007ni & 31.45990 & -0.33195 & 0.196 & 0.670 \\
19155 & 2007mn & 31.26644 & +0.17447 & 0.077 & 0.228 \\
19220 & 2007ox & 341.74185 & -0.07193 & 0.212 & 0.968 \\
19230 & 2007mo & 332.89093 & +0.76465 & 0.221 & 0.463 \\
19282 & 2007mk & 359.07211 & -0.50594 & 0.186 & 0.827 \\
19341 & 2007nf & 15.86018 & +0.33129 & 0.228 & 0.449 \\
19353 & 2007nj & 43.11432 & +0.25177 & 0.154 & 0.888 \\
19543 & 2007oj & 357.90829 & +0.27972 & 0.123 & 0.483 \\
19596 & 2007po & 53.88425 & +0.70331 & 0.292 & 0.996 \\
19604 & 2007oi & 5.32441 & +1.07446 & 0.296 & 0.097 \\
19616 & 2007ok & 37.10098 & +0.18458 & 0.166 & 0.963 \\
19625 & 2007pn & 34.14118 & -0.72247 & 0.307 & 0.534 \\
19632 & 2007ov & 40.28636 & +0.14436 & 0.315 & 0.951 \\
19658 & 2007ot & 8.90311 & -0.23275 & 0.200 & 0.993 \\
19702 & 2007pp & 47.75470 & +0.35671 & 0.262 & 0.822 \\
19775 & 2007pc & 318.95599 & +0.65120 & 0.138 & 0.936 \\
19794 & 2007oz & 359.31897 & +0.24923 & 0.297 & 0.517 \\
19818 & 2007pe & 35.26666 & +0.49634 & 0.304 & 0.779 \\
19913 & 2007qf & 333.76242 & -0.34134 & 0.204 & 0.212 \\
19940 & 2007pa & 315.39349 & -0.26855 & 0.157 & 0.991 \\
19968 & 2007ol & 24.34869 & -0.31209 & 0.056 & 0.904 \\
19969 & 2007pt & 31.91040 & -0.32408 & 0.175 & 0.972 \\
19990 & 2007ps & 34.80625 & -0.38490 & 0.246 & 0.833 \\
19992 & 2007pb & 357.10406 & -1.18509 & 0.228 & 0.938 \\
20039 & 2007qh & 9.87848 & +1.02425 & 0.248 & 0.899 \\
20040 & 2007rf & 328.87943 & +0.81499 & 0.288 & 0.655 \\
20048 & 2007pq & 339.30804 & +0.73627 & 0.185 & 0.804 \\
20064 & 2007om & 358.58612 & -0.91773 & 0.105 & 0.981 \\
20084 & 2007pd & 347.97513 & -0.57817 & 0.091 & 0.686 \\
20088 & \nodata & 13.20511 & +0.63137 & 0.244 & 0.466 \\
20097 & 2007rd & 311.75446 & -0.09939 & 0.221 & 0.992 \\
20106 & 2007pr & 346.55414 & +0.32884 & 0.333 & 0.922 \\
20111 & 2007pw & 354.39426 & +0.24802 & 0.245 & 0.396 \\
20142 & 2007qg & 23.00817 & -0.42981 & 0.314 & 0.997 \\
20184 & 2007qn & 359.78848 & +1.15823 & 0.324 & 0.824 \\
20186 & 2007pj & 357.29446 & +0.79778 & 0.354 & 0.927 \\
20245 & 2007pi & 341.70541 & +0.75618 & 0.288 & 0.976 \\
20345 & 2007qp & 10.70167 & +0.37962 & 0.265 & 0.316 \\
20350 & 2007ph & 312.80576 & -0.95590 & 0.130 & 0.082 \\
20364 & 2007qo & 25.75656 & -0.94553 & 0.218 & 0.671 \\
20376 & 2007re & 319.39539 & -0.52406 & 0.211 & 0.001 \\
20768 & 2007qq & 40.62584 & -0.97121 & 0.238 & 0.147 \\
20821 & 2007rk & 55.57257 & +1.06302 & 0.196 & 0.125 \\

\enddata
\end{deluxetable}

\begin{deluxetable}{rrrcc}
\tablecolumns{5}
\tabletypesize{\small}
\singlespace
\tablewidth{0pc} 
\tablecaption{SNe with Spectroscopic Redshifts in the Rate Sample
\label{tab:snerates2_105}
}
\tablehead{
\colhead{SN ID} & 
\colhead{$\alpha_{\mathrm{J2000}}$} &    
\colhead{$\delta_{\mathrm{J2000}}$} & 
\colhead{Redshift} &
\colhead{Fitprob} \\ 

\colhead{} & 
\colhead{[degrees]} &    
\colhead{[degrees]} & 
\colhead{} &
\colhead{}

}
\renewcommand{\arraystretch}{1.5}
\startdata
703 & 336.21786 & +0.65059 & 0.300 & 0.976 \\
779 & 26.67369 & -1.02072 & 0.238 & 0.986 \\
911 & 38.69067 & -0.11571 & 0.208 & 0.930 \\
1008 & 28.27810 & +1.11369 & 0.226 & 0.974 \\
1415 & 6.10647 & +0.59921 & 0.212 & 0.924 \\
1740 & 5.40428 & -0.88099 & 0.167 & 0.013 \\
2057 & 320.39969 & -0.31708 & 0.212 & 0.512 \\
2081 & 337.30505 & -1.20783 & 0.252 & 0.753 \\
2162 & 15.44242 & -0.13368 & 0.173 & 0.777 \\
2532 & 27.74742 & -0.23427 & 0.270 & 0.913 \\
2632 & 45.59013 & -1.22610 & 0.296 & 0.044 \\
2639 & 330.46411 & +0.66447 & 0.215 & 0.278 \\
2734 & 48.20686 & -0.69485 & 0.303 & 0.149 \\
2806 & 45.26690 & +0.27364 & 0.301 & 0.340 \\
2864 & 359.45090 & -1.23960 & 0.244 & 0.426 \\
3049 & 330.22342 & -1.23657 & 0.167 & 0.936 \\
3195 & 8.66830 & +0.24421 & 0.300 & 0.135 \\
3488 & 313.55518 & -1.01043 & 0.160 & 0.003 \\
3535 & 44.11722 & -0.13357 & 0.308 & 0.952 \\
3881 & 351.88162 & -0.44181 & 0.328 & 0.438 \\
4019 & 1.26167 & +1.14534 & 0.181 & 0.921 \\
4035 & 21.46727 & +1.04749 & 0.428 & 0.084 \\
4059 & 54.64996 & +0.14576 & 0.304 & 0.107 \\
4181 & 37.81778 & -1.13101 & 0.289 & 0.610 \\
4236 & 1.90560 & -1.01839 & 0.343 & 0.723 \\
4281 & 33.36735 & -0.96831 & 0.213 & 0.984 \\
4307 & 29.96247 & +0.94960 & 0.272 & 0.987 \\
4311 & 32.13080 & +1.01988 & 0.295 & 0.608 \\
4651 & 37.37555 & -0.74726 & 0.152 & 0.938 \\
4676 & 18.82365 & +0.78805 & 0.245 & 0.258 \\
4690 & 32.92946 & +0.68817 & 0.200 & 0.311 \\
5473 & 354.73663 & +0.38342 & 0.280 & 0.725 \\
5486 & 333.24792 & -0.41210 & 0.229 & 0.965 \\
5673 & 353.80124 & +0.78419 & 0.379 & 0.952 \\
5785 & 328.59738 & +0.08376 & 0.148 & 0.120 \\
5890 & 332.51556 & +0.60915 & 0.180 & 0.966 \\
5959 & 38.05976 & -0.30821 & 0.208 & 0.999 \\
5963 & 11.08100 & +0.47940 & 0.236 & 0.996 \\
5993 & 29.68238 & +0.04895 & 0.377 & 0.793 \\
6275 & 34.58972 & +0.02990 & 0.273 & 0.264 \\
6431 & 7.50847 & -0.70385 & 0.252 & 0.752 \\
6479 & 320.40723 & +0.58351 & 0.234 & 0.444 \\
6530 & 14.32909 & +0.02129 & 0.193 & 0.592 \\
6614 & 26.64692 & +0.86672 & 0.169 & 0.028 \\
6714 & 357.35992 & +0.63176 & 0.414 & 0.523 \\
6851 & 52.10445 & -0.04860 & 0.305 & 0.983 \\
6861 & 349.43219 & -1.11353 & 0.190 & 0.621 \\
6895 & 330.19290 & +0.92818 & 0.217 & 0.799 \\
6903 & 335.44351 & +0.97167 & 0.253 & 0.757 \\
7092 & 316.28043 & +1.22042 & 0.225 & 0.991 \\
7099 & 342.88809 & +1.16725 & 0.218 & 0.950 \\
7102 & 324.61981 & -0.61565 & 0.196 & 0.185 \\
7258 & 321.22992 & -0.99907 & 0.255 & 0.960 \\
7373 & 352.80481 & +0.58823 & 0.280 & 0.886 \\
7431 & 340.95438 & -0.27500 & 0.350 & 0.856 \\
7444 & 27.70291 & +0.42972 & 0.250 & 0.874 \\
7457 & 350.90311 & -0.37225 & 0.254 & 0.072 \\
7527 & 335.25632 & -1.20414 & 0.237 & 0.937 \\
7644 & 353.93500 & +0.86820 & 0.310 & 0.592 \\
7647 & 358.38300 & +1.04199 & 0.385 & 0.718 \\
7701 & 6.51694 & -1.22886 & 0.361 & 0.864 \\
7824 & 348.81302 & -0.08091 & 0.287 & 0.777 \\
7954 & 332.67694 & +0.34771 & 0.255 & 0.772 \\
8114 & 352.21658 & +0.22833 & 0.372 & 0.704 \\
8165 & 50.22528 & -1.10940 & 0.319 & 0.624 \\
8195 & 331.00635 & -0.89569 & 0.269 & 0.999 \\
8254 & 351.16376 & +0.81979 & 0.189 & 0.067 \\
8280 & 8.57363 & +0.79595 & 0.185 & 0.897 \\
8555 & 2.91543 & -0.41497 & 0.198 & 0.649 \\
8607 & 323.58691 & +0.15191 & 0.260 & 0.491 \\
8742 & 11.23943 & +0.43712 & 0.214 & 0.787 \\
9117 & 46.90165 & +0.98827 & 0.272 & 0.995 \\
9155 & 327.35437 & -0.78476 & 0.305 & 0.974 \\
9594 & 17.71423 & +0.12161 & 0.298 & 0.132 \\
9739 & 323.69379 & -0.87893 & 0.120 & 0.421 \\
11306 & 56.73846 & -0.51832 & 0.274 & 0.994 \\
12804 & 18.20149 & +1.04017 & 0.134 & 0.995 \\
12852 & 314.25305 & +0.68913 & 0.264 & 0.501 \\
13224 & 47.49494 & -0.24595 & 0.236 & 0.960 \\
13323 & 323.05118 & -0.13476 & 0.232 & 0.303 \\
13432 & 318.45322 & -1.07564 & 0.229 & 0.059 \\
13633 & 4.66560 & +0.00587 & 0.388 & 0.306 \\
13703 & 39.01363 & +1.25339 & 0.235 & 0.952 \\
13864 & 348.57407 & +0.19101 & 0.293 & 0.304 \\
13907 & 14.17929 & +0.23222 & 0.198 & 0.686 \\
13908 & 15.91596 & +0.29503 & 0.240 & 0.010 \\
13933 & 337.26547 & -0.56346 & 0.432 & 0.582 \\
13958 & 21.78776 & +0.80035 & 0.303 & 0.326 \\
14186 & 337.85046 & +0.99220 & 0.313 & 0.241 \\
14317 & 315.57065 & +0.33021 & 0.181 & 0.977 \\
14333 & 16.28526 & -0.01270 & 0.271 & 0.420 \\
14340 & 345.82657 & -0.85538 & 0.277 & 0.432 \\
14524 & 41.56276 & -0.79787 & 0.272 & 0.222 \\
14525 & 16.88274 & +0.47712 & 0.155 & 0.995 \\
14545 & 351.74817 & +1.01714 & 0.278 & 0.423 \\
14589 & 48.01208 & -0.67760 & 0.270 & 0.906 \\
14750 & 34.51896 & +0.65297 & 0.215 & 0.622 \\
14784 & 323.79834 & -0.34829 & 0.192 & 0.717 \\
14961 & 15.91944 & +0.93097 & 0.370 & 0.146 \\
15033 & 15.77839 & -0.15672 & 0.220 & 0.008 \\
15272 & 350.77301 & +0.08437 & 0.280 & 0.394 \\
15303 & 350.50882 & +0.54095 & 0.234 & 0.009 \\
15343 & 323.67233 & +0.68477 & 0.174 & 0.016 \\
15345 & 335.25677 & +0.81052 & 0.280 & 0.213 \\
15362 & 354.72998 & +0.76059 & 0.134 & 0.043 \\
15381 & 307.07623 & +0.49337 & 0.162 & 0.192 \\
15386 & 312.58475 & -1.23257 & 0.287 & 0.918 \\
15454 & 327.85934 & -0.84815 & 0.383 & 0.798 \\
15587 & 54.41718 & +0.99825 & 0.219 & 0.947 \\
15675 & 343.17361 & +0.36414 & 0.235 & 0.934 \\
15748 & 48.11442 & -0.13071 & 0.157 & 0.078 \\
15765 & 32.84710 & +0.24601 & 0.305 & 0.706 \\
15784 & 356.67365 & -0.61540 & 0.277 & 0.794 \\
15802 & 29.75320 & -1.07942 & 0.342 & 0.999 \\
15806 & 24.09219 & -0.83070 & 0.250 & 0.569 \\
15823 & 314.25256 & +0.19901 & 0.215 & 0.989 \\
15850 & 0.66751 & -1.16522 & 0.250 & 0.461 \\
15892 & 323.19894 & +0.68938 & 0.184 & 0.194 \\
15909 & 11.31475 & +0.79673 & 0.218 & 0.010 \\
15971 & 40.11259 & +0.52619 & 0.316 & 0.056 \\
16052 & 58.60019 & -0.72081 & 0.144 & 0.923 \\
16091 & 50.19004 & +0.84181 & 0.301 & 0.996 \\
16103 & 312.97516 & -1.05017 & 0.202 & 0.069 \\
16152 & 46.90844 & +0.98769 & 0.272 & 0.715 \\
16163 & 31.49918 & -0.85579 & 0.155 & 0.204 \\
16462 & 17.04058 & -0.38646 & 0.245 & 0.486 \\
16467 & 328.58572 & +0.11822 & 0.221 & 0.062 \\
16768 & 322.70056 & +0.69283 & 0.169 & 0.084 \\
17206 & 45.98524 & +0.72816 & 0.156 & 0.301 \\
17434 & 18.43727 & -0.07290 & 0.179 & 0.840 \\
17748 & 9.80709 & -0.27714 & 0.179 & 0.971 \\
17882 & 15.01838 & +1.24641 & 0.270 & 0.102 \\
17928 & 358.47009 & +1.11463 & 0.197 & 0.975 \\
17958 & 34.43497 & -0.71299 & 0.276 & 0.377 \\
18047 & 22.07419 & -0.65805 & 0.359 & 0.025 \\
18630 & 347.97998 & -0.26443 & 0.359 & 0.009 \\
18647 & 322.89191 & -0.30387 & 0.213 & 0.389 \\
19525 & 23.78299 & -0.18406 & 0.153 & 0.388 \\
19787 & 0.28111 & -0.09813 & 0.197 & 0.495 \\
19833 & 38.87204 & -1.17885 & 0.233 & 0.034 \\
20141 & 357.54199 & -0.52391 & 0.342 & 0.647 \\
20171 & 3.01303 & +0.21357 & 0.240 & 0.034 \\
20232 & 7.08330 & -0.05822 & 0.217 & 0.974 \\
20626 & 8.47575 & -0.59308 & 0.276 & 0.313 \\
20721 & 323.18481 & -0.62219 & 0.212 & 0.180 \\
20788 & 51.67339 & -0.47783 & 0.394 & 0.954 \\
20896 & 0.72633 & +0.93933 & 0.361 & 0.740 \\
21024 & 43.94143 & -0.05122 & 0.262 & 0.005 \\

\enddata
\end{deluxetable}

\begin{deluxetable}{rrrccc}
\tablecolumns{6}
\tabletypesize{\small}
\singlespace
\tablewidth{0pc} 
\tablecaption{SNe with Photometric Redshifts in the Rate Sample
\label{tab:snerates2_photoz}
}
\tablehead{
\colhead{SN ID} & 
\colhead{$\alpha_{\mathrm{J2000}}$} &    
\colhead{$\delta_{\mathrm{J2000}}$} & 
\colhead{Redshift} &
\colhead{Redshift Error} &
\colhead{Fitprob} \\ 

\colhead{} & 
\colhead{[degrees]} &    
\colhead{[degrees]} & 
\colhead{} &
\colhead{}
}
\renewcommand{\arraystretch}{1.5}
\startdata
822 & 40.56070 & -0.86217 & 0.220 & 0.018 & 0.090 \\
1342 & 346.52740 & +0.11688 & 0.283 & 0.026 & 0.962 \\
1403 & 359.70386 & +0.43188 & 0.341 & 0.034 & 0.112 \\
1658 & 357.50443 & +0.65006 & 0.256 & 0.025 & 0.836 \\
1899 & 323.34305 & -0.70642 & 0.341 & 0.037 & 0.005 \\
2784 & 28.07526 & -0.04169 & 0.381 & 0.021 & 0.488 \\
2855 & 16.17518 & -0.35642 & 0.233 & 0.016 & 0.978 \\
3206 & 13.57741 & +0.41816 & 0.387 & 0.031 & 0.003 \\
3368 & 44.45594 & +1.23082 & 0.320 & 0.035 & 0.831 \\
3417 & 314.27606 & +0.97825 & 0.262 & 0.016 & 0.413 \\
3506 & 336.25064 & -0.97821 & 0.211 & 0.011 & 0.963 \\
3945 & 346.00906 & -0.28307 & 0.260 & 0.017 & 0.070 \\
3975 & 29.82097 & +0.20364 & 0.399 & 0.019 & 0.002 \\
3983 & 7.27582 & -0.25663 & 0.279 & 0.021 & 0.220 \\
4028 & 11.01385 & +1.24247 & 0.327 & 0.030 & 0.724 \\
4044 & 33.41526 & +1.24086 & 0.383 & 0.027 & 0.395 \\
4079 & 29.20663 & +0.75403 & 0.414 & 0.015 & 0.078 \\
4360 & 19.65592 & +0.91577 & 0.328 & 0.021 & 0.932 \\
4455 & 52.15621 & -0.06056 & 0.345 & 0.040 & 0.001 \\
4558 & 29.88546 & +0.29007 & 0.344 & 0.043 & 0.501 \\
4572 & 36.26846 & +0.39601 & 0.386 & 0.025 & 0.058 \\
4578 & 38.59660 & +0.31129 & 0.341 & 0.033 & 0.516 \\
4579 & 38.67472 & +0.36412 & 0.406 & 0.024 & 0.884 \\
4714 & 37.15236 & -0.20551 & 0.435 & 0.011 & 0.016 \\
4757 & 34.14671 & -1.01575 & 0.416 & 0.021 & 0.232 \\
4803 & 32.56401 & +0.30202 & 0.398 & 0.022 & 0.636 \\
5199 & 348.79242 & -0.99494 & 0.248 & 0.021 & 0.761 \\
5235 & 337.22424 & +0.63615 & 0.253 & 0.028 & 0.373 \\
5468 & 342.53067 & +0.40476 & 0.277 & 0.017 & 0.965 \\
5524 & 316.57864 & -0.99036 & 0.299 & 0.024 & 0.175 \\
5543 & 356.85907 & +0.30325 & 0.330 & 0.022 & 0.808 \\
5702 & 12.58325 & -0.91914 & 0.222 & 0.009 & 0.931 \\
5731 & 11.33848 & -0.54451 & 0.376 & 0.027 & 0.300 \\
5735 & 311.65878 & +0.65057 & 0.229 & 0.019 & 0.785 \\
5792 & 328.71417 & -1.23786 & 0.226 & 0.045 & 0.066 \\
5802 & 328.14502 & +0.84568 & 0.270 & 0.020 & 0.996 \\
5803 & 335.97989 & +0.93223 & 0.285 & 0.023 & 0.360 \\
5917 & 6.03683 & -0.25513 & 0.254 & 0.019 & 0.994 \\
6055 & 37.90630 & -0.89602 & 0.404 & 0.020 & 0.250 \\
6225 & 0.28751 & -1.00040 & 0.333 & 0.026 & 0.225 \\
6501 & 315.15686 & +0.13145 & 0.338 & 0.024 & 0.850 \\
6560 & 321.44653 & +0.84973 & 0.290 & 0.031 & 0.264 \\
6618 & 41.91954 & +0.96416 & 0.313 & 0.015 & 1.000 \\
6877 & 52.50651 & +0.62126 & 0.321 & 0.013 & 0.003 \\
6889 & 326.71991 & +0.90946 & 0.290 & 0.032 & 0.980 \\
6912 & 315.73236 & -0.64258 & 0.369 & 0.041 & 0.991 \\
7205 & 44.10721 & +0.69505 & 0.315 & 0.014 & 0.799 \\
7304 & 24.83592 & -0.65620 & 0.256 & 0.013 & 0.932 \\
7357 & 321.63586 & -0.36266 & 0.392 & 0.029 & 0.633 \\
7365 & 352.31717 & +0.57975 & 0.406 & 0.023 & 0.911 \\
7413 & 15.04521 & +0.47463 & 0.380 & 0.031 & 0.949 \\
7636 & 320.85507 & +0.11879 & 0.396 & 0.021 & 0.540 \\
7654 & 2.47238 & +0.96446 & 0.388 & 0.019 & 0.997 \\
7656 & 7.75481 & +1.01067 & 0.335 & 0.046 & 0.057 \\
7699 & 6.23865 & -1.21577 & 0.355 & 0.024 & 0.336 \\
7712 & 10.95986 & -1.24082 & 0.426 & 0.015 & 0.587 \\
7717 & 14.44051 & -1.23264 & 0.329 & 0.025 & 0.741 \\
7802 & 345.57590 & +0.72910 & 0.310 & 0.026 & 0.650 \\
7803 & 346.08633 & +0.69040 & 0.382 & 0.032 & 0.176 \\
7841 & 10.31183 & -0.05151 & 0.409 & 0.022 & 0.663 \\
7884 & 21.76736 & +0.12340 & 0.344 & 0.035 & 0.991 \\
8092 & 50.15116 & +0.09387 & 0.329 & 0.022 & 0.696 \\
8118 & 355.85870 & +0.29712 & 0.415 & 0.025 & 0.658 \\
8138 & 11.46659 & +0.37805 & 0.331 & 0.027 & 0.816 \\
8226 & 334.81720 & +0.74912 & 0.429 & 0.015 & 0.654 \\
8297 & 24.97450 & +0.69145 & 0.255 & 0.019 & 0.799 \\
8700 & 35.10958 & +0.22706 & 0.385 & 0.031 & 0.510 \\
8705 & 40.43310 & +0.26872 & 0.384 & 0.034 & 0.336 \\
8793 & 11.93775 & -0.56470 & 0.383 & 0.034 & 0.023 \\
9052 & 21.04114 & -0.48308 & 0.237 & 0.017 & 0.887 \\
9109 & 26.38895 & +0.85063 & 0.259 & 0.018 & 0.921 \\
9132 & 16.25518 & +0.54001 & 0.368 & 0.034 & 0.715 \\
9218 & 46.70148 & -0.70051 & 0.236 & 0.020 & 0.060 \\
9334 & 346.99301 & +0.84164 & 0.336 & 0.036 & 0.242 \\
9632 & 30.17027 & +1.24947 & 0.383 & 0.016 & 0.691 \\
9895 & 35.26586 & +0.50430 & 0.328 & 0.027 & 0.385 \\
10113 & 54.18169 & -0.14042 & 0.296 & 0.022 & 0.985 \\
10559 & 354.11643 & -1.22660 & 0.257 & 0.027 & 0.696 \\
11092 & 302.62335 & +1.12162 & 0.090 & 0.005 & 0.711 \\
12936 & 32.04546 & -1.05151 & 0.194 & 0.029 & 0.567 \\
13015 & 25.30317 & +0.92776 & 0.225 & 0.008 & 0.002 \\
13016 & 25.58846 & +0.97958 & 0.244 & 0.018 & 0.749 \\
13064 & 343.67349 & -1.13961 & 0.228 & 0.020 & 0.943 \\
13073 & 336.02512 & +0.14211 & 0.323 & 0.035 & 0.882 \\
13096 & 358.37015 & -1.20224 & 0.363 & 0.032 & 0.508 \\
13098 & 359.17716 & -1.19228 & 0.276 & 0.019 & 0.474 \\
13108 & 9.87074 & -1.20040 & 0.288 & 0.034 & 0.136 \\
13144 & 20.08210 & -0.33359 & 0.275 & 0.016 & 0.012 \\
13147 & 358.19369 & +0.08943 & 0.370 & 0.032 & 0.496 \\
13168 & 1.60157 & +0.62138 & 0.337 & 0.029 & 0.699 \\
13329 & 340.72012 & -0.10852 & 0.319 & 0.022 & 0.848 \\
13334 & 321.67703 & +0.39447 & 0.364 & 0.030 & 0.201 \\
13437 & 328.74725 & -0.66195 & 0.379 & 0.024 & 0.980 \\
13441 & 346.95294 & -0.23305 & 0.282 & 0.035 & 0.741 \\
13460 & 345.71448 & +0.14190 & 0.299 & 0.026 & 0.011 \\
13474 & 11.82052 & -1.06794 & 0.328 & 0.013 & 0.021 \\
13476 & 356.01660 & -0.79097 & 0.310 & 0.028 & 0.383 \\
13477 & 7.80524 & -0.70838 & 0.407 & 0.022 & 0.182 \\
13491 & 43.28595 & +0.11893 & 0.372 & 0.016 & 0.008 \\
13554 & 7.15682 & +0.87184 & 0.332 & 0.028 & 0.719 \\
13615 & 311.41235 & +1.18661 & 0.232 & 0.021 & 0.854 \\
13646 & 357.47256 & -1.03215 & 0.292 & 0.016 & 0.608 \\
13649 & 355.52835 & -0.61896 & 0.319 & 0.027 & 0.493 \\
13675 & 310.65897 & -0.91660 & 0.292 & 0.022 & 0.225 \\
13729 & 318.91867 & +0.87517 & 0.343 & 0.041 & 0.741 \\
13737 & 338.92618 & +0.96260 & 0.329 & 0.029 & 0.060 \\
13740 & 311.21704 & -1.08869 & 0.241 & 0.018 & 0.027 \\
13768 & 323.07596 & -0.76319 & 0.240 & 0.016 & 0.808 \\
13813 & 318.32089 & -0.40486 & 0.240 & 0.016 & 0.994 \\
13833 & 354.32144 & -0.20943 & 0.419 & 0.019 & 0.209 \\
13843 & 319.16034 & +0.20718 & 0.433 & 0.012 & 0.015 \\
13859 & 11.06211 & +0.96698 & 0.324 & 0.021 & 0.961 \\
13861 & 337.84750 & +0.07537 & 0.365 & 0.028 & 0.684 \\
13867 & 351.45483 & +0.02935 & 0.333 & 0.038 & 0.538 \\
13896 & 2.71280 & -0.06992 & 0.229 & 0.015 & 0.538 \\
13901 & 21.92091 & -0.14467 & 0.421 & 0.017 & 0.001 \\
13904 & 6.85967 & +0.28349 & 0.244 & 0.015 & 0.032 \\
13909 & 16.41927 & +0.31172 & 0.293 & 0.011 & 0.317 \\
13952 & 4.63440 & +0.78849 & 0.369 & 0.041 & 0.133 \\
14074 & 342.12595 & -0.82529 & 0.321 & 0.028 & 0.771 \\
14093 & 26.53920 & -1.05637 & 0.427 & 0.015 & 0.205 \\
14206 & 17.39464 & +0.70405 & 0.237 & 0.012 & 0.579 \\
14231 & 57.64965 & +0.78754 & 0.162 & 0.013 & 0.964 \\
14268 & 52.19934 & +0.39107 & 0.259 & 0.013 & 0.870 \\
14304 & 311.00174 & -0.60592 & 0.322 & 0.022 & 0.771 \\
14322 & 353.77292 & -0.16948 & 0.377 & 0.030 & 0.603 \\
14343 & 0.70710 & -0.95408 & 0.274 & 0.025 & 0.665 \\
14347 & 16.98071 & -1.01713 & 0.296 & 0.021 & 0.348 \\
14357 & 47.64600 & -0.93812 & 0.414 & 0.019 & 0.183 \\
14361 & 43.51542 & -0.10534 & 0.375 & 0.016 & 0.950 \\
14372 & 50.81182 & -0.14131 & 0.301 & 0.022 & 0.778 \\
14403 & 16.83825 & +0.65173 & 0.293 & 0.012 & 0.114 \\
14438 & 340.06516 & -1.06933 & 0.282 & 0.011 & 0.470 \\
14444 & 336.70151 & -0.81572 & 0.244 & 0.020 & 0.339 \\
14453 & 316.28625 & +0.94936 & 0.349 & 0.028 & 0.305 \\
14463 & 3.14527 & -0.35054 & 0.207 & 0.019 & 0.981 \\
14467 & 25.60195 & -0.31297 & 0.358 & 0.020 & 0.810 \\
14470 & 44.21490 & -0.35178 & 0.179 & 0.010 & 0.998 \\
14522 & 28.52871 & -0.72565 & 0.389 & 0.036 & 0.828 \\
14528 & 349.93512 & +0.47960 & 0.337 & 0.031 & 0.543 \\
14531 & 2.62424 & +0.48453 & 0.356 & 0.023 & 0.025 \\
14539 & 28.24310 & +0.50526 & 0.379 & 0.029 & 0.528 \\
14540 & 36.51231 & +0.58206 & 0.257 & 0.011 & 0.999 \\
14549 & 8.18106 & +0.96332 & 0.229 & 0.022 & 0.725 \\
14561 & 46.69879 & +0.94942 & 0.214 & 0.016 & 0.018 \\
14588 & 47.68921 & -0.77679 & 0.372 & 0.029 & 0.685 \\
14617 & 53.36466 & +1.02023 & 0.258 & 0.015 & 0.796 \\
14638 & 358.55978 & -0.12483 & 0.331 & 0.023 & 0.366 \\
14644 & 24.08068 & -0.06971 & 0.405 & 0.030 & 0.253 \\
14708 & 15.31452 & -0.46494 & 0.337 & 0.032 & 0.701 \\
14760 & 56.25041 & +0.72820 & 0.321 & 0.025 & 0.954 \\
14763 & 33.56654 & +1.12131 & 0.351 & 0.041 & 0.664 \\
14786 & 331.82040 & -0.32145 & 0.321 & 0.034 & 0.897 \\
14809 & 331.37582 & -0.65592 & 0.311 & 0.026 & 0.157 \\
14823 & 20.80287 & -0.31058 & 0.279 & 0.035 & 0.815 \\
14888 & 25.70452 & -0.64986 & 0.366 & 0.014 & 0.699 \\
14900 & 45.23616 & -0.63561 & 0.432 & 0.012 & 0.100 \\
14951 & 53.00833 & +0.41389 & 0.427 & 0.014 & 0.023 \\
14965 & 18.93550 & +1.03783 & 0.238 & 0.028 & 0.793 \\
15055 & 58.49916 & -0.05169 & 0.196 & 0.015 & 1.000 \\
15103 & 338.09860 & -0.50470 & 0.374 & 0.031 & 0.715 \\
15108 & 24.43465 & -0.54161 & 0.412 & 0.018 & 0.213 \\
15137 & 356.10730 & -0.80658 & 0.281 & 0.021 & 0.899 \\
15198 & 34.62743 & -0.21229 & 0.290 & 0.008 & 1.000 \\
15260 & 339.17731 & -0.27577 & 0.215 & 0.016 & 0.983 \\
15263 & 325.80643 & +0.00996 & 0.389 & 0.031 & 0.523 \\
15268 & 338.44495 & +0.13868 & 0.275 & 0.019 & 0.758 \\
15289 & 334.43304 & -1.19865 & 0.376 & 0.022 & 0.818 \\
15291 & 338.08383 & -1.15178 & 0.313 & 0.030 & 0.336 \\
15294 & 318.47522 & -0.73191 & 0.347 & 0.046 & 0.115 \\
15324 & 31.69360 & -0.83908 & 0.388 & 0.017 & 0.001 \\
15351 & 326.53534 & -0.05332 & 0.326 & 0.047 & 0.831 \\
15357 & 335.33731 & +0.23419 & 0.266 & 0.026 & 0.646 \\
15359 & 347.45624 & +0.31050 & 0.262 & 0.018 & 0.071 \\
15363 & 13.03311 & +0.66130 & 0.356 & 0.038 & 0.331 \\
15417 & 15.55787 & +0.46188 & 0.304 & 0.027 & 0.859 \\
15419 & 19.90523 & +0.88563 & 0.267 & 0.021 & 0.937 \\
15423 & 36.50804 & +0.62611 & 0.318 & 0.014 & 0.134 \\
15428 & 39.09415 & +1.04246 & 0.410 & 0.014 & 0.099 \\
15448 & 52.38031 & -1.14735 & 0.232 & 0.016 & 0.936 \\
15455 & 330.08044 & -1.05142 & 0.395 & 0.030 & 0.389 \\
15483 & 319.79196 & +0.69697 & 0.321 & 0.033 & 0.901 \\
15489 & 7.75260 & -0.07103 & 0.365 & 0.028 & 0.996 \\
15496 & 8.07390 & +0.27446 & 0.237 & 0.011 & 0.949 \\
15511 & 44.93024 & -0.95519 & 0.256 & 0.026 & 0.993 \\
15525 & 51.52761 & +1.13072 & 0.355 & 0.045 & 0.729 \\
15553 & 321.12012 & +0.94418 & 0.285 & 0.035 & 0.935 \\
15592 & 320.46732 & -1.17850 & 0.341 & 0.030 & 0.090 \\
15710 & 55.05272 & +0.72199 & 0.363 & 0.028 & 0.609 \\
15719 & 37.64792 & +1.10330 & 0.249 & 0.026 & 0.612 \\
15726 & 349.49463 & -0.00728 & 0.394 & 0.017 & 0.124 \\
15745 & 42.20790 & -0.10763 & 0.414 & 0.022 & 0.941 \\
15751 & 54.02817 & -0.05176 & 0.331 & 0.035 & 0.784 \\
15777 & 326.18854 & -0.61831 & 0.296 & 0.013 & 0.405 \\
15782 & 352.67618 & -0.42804 & 0.314 & 0.031 & 0.058 \\
15785 & 30.27846 & -0.53928 & 0.331 & 0.024 & 0.172 \\
15812 & 19.35990 & -0.30902 & 0.312 & 0.023 & 0.300 \\
15814 & 29.72514 & -0.37273 & 0.350 & 0.021 & 0.971 \\
15816 & 19.45456 & +0.09876 & 0.326 & 0.043 & 0.110 \\
15817 & 29.56453 & +0.01327 & 0.394 & 0.026 & 0.306 \\
15829 & 323.08551 & -0.76231 & 0.322 & 0.025 & 0.967 \\
15860 & 4.59735 & +0.18303 & 0.288 & 0.016 & 0.959 \\
15870 & 30.01619 & -0.23499 & 0.396 & 0.025 & 0.439 \\
15874 & 39.76944 & -0.38710 & 0.410 & 0.026 & 0.645 \\
15903 & 48.90611 & -0.60267 & 0.283 & 0.060 & 0.047 \\
15994 & 11.93237 & +0.10035 & 0.243 & 0.040 & 0.168 \\
16111 & 329.31693 & +0.93545 & 0.230 & 0.019 & 0.999 \\
16130 & 51.77484 & +0.15773 & 0.325 & 0.020 & 0.171 \\
16148 & 48.66113 & +0.58814 & 0.308 & 0.024 & 0.916 \\
16199 & 332.94424 & +1.13495 & 0.232 & 0.033 & 0.575 \\
16220 & 9.92570 & +0.68008 & 0.314 & 0.038 & 0.625 \\
16225 & 36.51204 & +0.26114 & 0.390 & 0.025 & 0.555 \\
16237 & 23.01032 & -0.62387 & 0.296 & 0.015 & 0.010 \\
16238 & 27.91512 & -0.43587 & 0.339 & 0.043 & 0.466 \\
16247 & 56.36977 & -0.46345 & 0.311 & 0.052 & 0.110 \\
16302 & 331.76721 & +0.18343 & 0.187 & 0.013 & 0.003 \\
16410 & 319.71500 & +0.64812 & 0.309 & 0.020 & 0.051 \\
16460 & 11.00961 & -0.23004 & 0.286 & 0.018 & 0.829 \\
17541 & 325.05933 & +0.72586 & 0.261 & 0.026 & 0.628 \\
17577 & 331.91479 & -1.05463 & 0.283 & 0.035 & 0.015 \\
17647 & 34.28204 & -0.80163 & 0.249 & 0.012 & 0.922 \\
17695 & 41.98984 & +0.58951 & 0.170 & 0.014 & 0.337 \\
17773 & 33.28418 & -0.30634 & 0.273 & 0.021 & 0.790 \\
17820 & 16.13956 & -0.46653 & 0.306 & 0.021 & 0.544 \\
17829 & 53.03160 & -0.83546 & 0.332 & 0.037 & 0.133 \\
17899 & 353.69949 & -0.15169 & 0.252 & 0.027 & 0.935 \\
17906 & 56.23551 & +0.40241 & 0.184 & 0.012 & 0.995 \\
17925 & 346.30588 & +0.30397 & 0.286 & 0.020 & 0.444 \\
17949 & 29.20499 & -0.23557 & 0.305 & 0.020 & 0.358 \\
17965 & 340.82355 & -0.75888 & 0.296 & 0.034 & 0.986 \\
18041 & 10.89522 & -0.69872 & 0.400 & 0.021 & 0.246 \\
18049 & 6.31049 & -0.36792 & 0.261 & 0.021 & 0.411 \\
18083 & 36.34912 & -1.20823 & 0.179 & 0.029 & 0.541 \\
18146 & 32.42931 & -0.00044 & 0.432 & 0.016 & 0.370 \\
18243 & 335.04486 & -0.74009 & 0.278 & 0.023 & 0.998 \\
18253 & 345.58429 & -0.90850 & 0.252 & 0.049 & 0.020 \\
18283 & 39.75129 & -0.54572 & 0.232 & 0.038 & 0.484 \\
18324 & 6.38100 & +0.79376 & 0.278 & 0.036 & 0.958 \\
18339 & 31.87198 & +1.15926 & 0.303 & 0.021 & 0.670 \\
18362 & 10.13653 & -0.18209 & 0.220 & 0.028 & 0.956 \\
18374 & 2.56478 & -0.17714 & 0.286 & 0.015 & 0.999 \\
18405 & 313.54889 & -0.91769 & 0.236 & 0.025 & 0.950 \\
18479 & 347.95078 & -0.81114 & 0.290 & 0.018 & 0.571 \\
18582 & 308.66626 & -0.63135 & 0.266 & 0.015 & 0.718 \\
18588 & 338.47177 & -1.20459 & 0.348 & 0.031 & 0.559 \\
18589 & 340.83459 & -1.14140 & 0.334 & 0.018 & 0.130 \\
18651 & 332.45709 & +0.11260 & 0.214 & 0.029 & 0.875 \\
18666 & 342.56442 & +0.19474 & 0.301 & 0.016 & 0.992 \\
18704 & 11.52182 & -0.52672 & 0.302 & 0.025 & 0.557 \\
18801 & 49.76012 & -0.45803 & 0.379 & 0.041 & 0.770 \\
18824 & 23.40139 & +1.19070 & 0.388 & 0.019 & 0.428 \\
18884 & 326.26300 & -0.94996 & 0.155 & 0.017 & 0.001 \\
18920 & 53.22518 & +0.53269 & 0.279 & 0.027 & 0.588 \\
18942 & 17.49710 & -0.01868 & 0.432 & 0.015 & 0.379 \\
18943 & 20.76962 & -0.06722 & 0.328 & 0.036 & 0.762 \\
18947 & 11.18694 & -0.87064 & 0.329 & 0.029 & 0.306 \\
18952 & 11.30682 & +0.82777 & 0.313 & 0.041 & 0.152 \\
18971 & 31.44543 & +1.20877 & 0.263 & 0.028 & 0.648 \\
18990 & 15.97268 & -0.59233 & 0.407 & 0.025 & 0.726 \\
18993 & 24.32056 & -0.50393 & 0.303 & 0.029 & 0.299 \\
19000 & 37.95206 & -0.49290 & 0.271 & 0.010 & 0.377 \\
19001 & 41.64711 & -0.45204 & 0.271 & 0.024 & 0.254 \\
19205 & 322.18607 & -0.84680 & 0.313 & 0.022 & 0.973 \\
19209 & 344.26331 & -0.88774 & 0.163 & 0.018 & 0.061 \\
19274 & 353.88544 & -0.57059 & 0.328 & 0.027 & 0.798 \\
19322 & 355.44455 & +1.23752 & 0.311 & 0.014 & 0.200 \\
19335 & 350.05820 & +0.39510 & 0.323 & 0.046 & 0.496 \\
19336 & 358.87741 & +0.26392 & 0.386 & 0.017 & 0.052 \\
19339 & 5.99431 & +0.31237 & 0.319 & 0.020 & 0.536 \\
19347 & 30.75784 & +0.35759 & 0.294 & 0.016 & 0.982 \\
19352 & 40.32961 & +0.25313 & 0.334 & 0.033 & 0.569 \\
19399 & 55.71893 & +0.83188 & 0.251 & 0.020 & 0.881 \\
19459 & 47.33323 & -0.73490 & 0.136 & 0.009 & 0.633 \\
19467 & 340.96490 & -0.97697 & 0.302 & 0.036 & 0.710 \\
19545 & 330.44305 & +0.40112 & 0.243 & 0.032 & 0.167 \\
19556 & 17.70001 & +0.22681 & 0.344 & 0.018 & 0.923 \\
19593 & 29.96931 & +0.67895 & 0.307 & 0.034 & 0.227 \\
19708 & 42.17281 & +0.65459 & 0.165 & 0.010 & 0.667 \\
19723 & 338.89166 & +0.41763 & 0.260 & 0.024 & 0.802 \\
19769 & 350.44351 & -0.97642 & 0.200 & 0.020 & 0.276 \\
19777 & 349.08566 & -0.50685 & 0.421 & 0.022 & 0.020 \\
19804 & 12.21571 & -0.21600 & 0.320 & 0.016 & 0.008 \\
19821 & 356.77942 & +1.02593 & 0.245 & 0.026 & 0.736 \\
19825 & 34.63781 & +0.88079 & 0.165 & 0.014 & 0.928 \\
19848 & 49.42086 & +0.90219 & 0.262 & 0.033 & 0.583 \\
19986 & 29.62428 & +0.15794 & 0.363 & 0.028 & 0.947 \\
19987 & 16.04479 & -0.32652 & 0.255 & 0.015 & 0.020 \\
20033 & 22.93019 & -0.73191 & 0.216 & 0.021 & 0.901 \\
20046 & 321.49316 & +0.80122 & 0.261 & 0.019 & 0.940 \\
20090 & 301.89996 & -0.07334 & 0.201 & 0.018 & 0.790 \\
20104 & 318.13837 & +0.28148 & 0.314 & 0.022 & 0.015 \\
20231 & 4.58886 & -0.03239 & 0.374 & 0.016 & 0.700 \\
20272 & 17.21835 & -0.02123 & 0.414 & 0.023 & 0.659 \\
20276 & 34.56314 & -0.11415 & 0.359 & 0.024 & 0.709 \\
20278 & 36.89183 & -0.18167 & 0.325 & 0.024 & 0.058 \\
20467 & 20.01988 & +0.91342 & 0.277 & 0.022 & 0.998 \\
20476 & 325.42004 & +0.83897 & 0.304 & 0.016 & 0.733 \\
20491 & 350.98157 & -1.19331 & 0.228 & 0.019 & 0.999 \\
20497 & 11.45800 & -1.18681 & 0.333 & 0.026 & 0.545 \\
20514 & 23.07604 & -0.75532 & 0.388 & 0.022 & 0.026 \\
20534 & 339.60840 & -0.23645 & 0.286 & 0.010 & 0.117 \\
20537 & 314.83130 & -0.30518 & 0.338 & 0.022 & 0.322 \\
20612 & 319.46490 & -0.47161 & 0.343 & 0.037 & 0.484 \\
20663 & 16.45225 & +0.82060 & 0.295 & 0.026 & 0.898 \\
20722 & 325.25613 & +0.70129 & 0.295 & 0.025 & 0.572 \\
20744 & 346.54922 & +0.35883 & 0.233 & 0.021 & 0.805 \\
20750 & 23.14062 & +0.41276 & 0.291 & 0.027 & 0.903 \\
20791 & 317.94785 & -0.38600 & 0.319 & 0.039 & 0.534 \\
20819 & 43.20791 & +1.21915 & 0.318 & 0.020 & 0.083 \\
20844 & 13.00876 & -1.12150 & 0.406 & 0.024 & 0.512 \\
20971 & 29.04193 & -0.21597 & 0.293 & 0.019 & 0.235 \\
21015 & 38.37030 & -0.98158 & 0.376 & 0.029 & 0.843 \\
21081 & 12.03145 & -0.55571 & 0.323 & 0.032 & 0.090 \\
21306 & 352.92325 & +0.09141 & 0.315 & 0.034 & 0.976 \\

\enddata
\end{deluxetable}

\begin{deluxetable}{ccrr}
\tablecolumns{4}
\tabletypesize{\small}
\singlespace
\tablewidth{0pc}
\tablecaption{SN Rate vs.~Redshift
\label{tab:midzratez1}
}
\tablehead{
\colhead{Redshift} & 
\colhead{SN Rate\tablenotemark{a}} &
\colhead{$-\Delta N/N$\tablenotemark{b}} &
\colhead{$N_{\mathrm{CC}}/N_{\mathrm{Ia}}$} \\

\colhead{} &
\colhead{[$10^{-5}$ SNe $\mathrm{yr}^{-1} ~\mathrm{Mpc}^{-3} ~h^{3}_{70}]$}  &
\colhead{} 
}

\renewcommand{\arraystretch}{1.5}
\startdata
0.025 - 0.050 & $2.78^{+1.12+0.15}_{-0.83-0.00}$ &  0.00 \% & \nodata \\
0.075 - 0.125 & $2.59^{+0.52+0.18}_{-0.44-0.01}$ & -0.06 \% & $0.71^{+0.56}_{-0.33} $ \% \\
0.125 - 0.175 & $3.07^{+0.38+0.35}_{-0.34-0.05}$ & -0.21 \% & $2.99^{+2.37}_{-1.40} $ \% \\
0.175 - 0.225 & $3.48^{+0.32+0.82}_{-0.30-0.07}$ & -0.21 \% & $2.71^{+2.14}_{-1.27} $ \% \\
0.225 - 0.275 & $3.65^{+0.31+1.82}_{-0.28-0.12}$ & +0.28 \% & $2.06^{+1.63}_{-0.96} $ \% \\
0.275 - 0.325 & $4.34^{+0.37+3.96}_{-0.34-0.16}$ & +1.86 \% & $0.67^{+0.53}_{-0.31} $ \% \\
\enddata

\tablenotetext{a}{The errors given are statistical and systematic, respectively.}
\tablenotetext{b}{Assuming a rate model $r_V \propto (1+z)^{\nuval}$; see \S \ref{sec:dndzbias}.}
\end{deluxetable}

\acknowledgements

Funding for the SDSS and SDSS-II has been provided by the Alfred
P. Sloan Foundation, the Participating Institutions, the National
Science Foundation (NSF), the U.S. Department of Energy, the National
Aeronautics and Space Administration (NASA), the Japanese Monbukagakusho, the
Max Planck Society, and the Higher Education Funding Council for
England. The SDSS Web Site is http://www.sdss.org/.

The SDSS is managed by the Astrophysical Research Consortium for the
Participating Institutions. The Participating Institutions are the
American Museum of Natural History, Astrophysical Institute Potsdam,
University of Basel, University of Cambridge, Case Western Reserve
University, University of Chicago, Drexel University, Fermilab, the
Institute for Advanced Study, the Japan Participation Group, Johns
Hopkins University, the Joint Institute for Nuclear Astrophysics, the
Kavli Institute for Particle Astrophysics and Cosmology, the Korean
Scientist Group, the Chinese Academy of Sciences (LAMOST), Los Alamos
National Laboratory, the Max-Planck-Institute for Astronomy (MPIA),
the Max-Planck-Institute for Astrophysics (MPA), New Mexico State
University, Ohio State University, University of Pittsburgh,
University of Portsmouth, Princeton University, the United States
Naval Observatory, and the University of Washington.

This work is based in part on observations made at the following
telescopes.  The Hobby-Eberly Telescope (HET) is a joint project of
the University of Texas at Austin, the Pennsylvania State University,
Stanford University, Ludwig-Maximillians-Universit\"at M\"unchen, and
Georg-August-Universit\"at G\"ottingen.  The HET is named in honor of
its principal benefactors, William P. Hobby and Robert E. Eberly.  The
Marcario Low-Resolution Spectrograph is named for Mike Marcario of
High Lonesome Optics, who fabricated several optical elements for the
instrument but died before its completion; it is a joint project of
the Hobby-Eberly Telescope partnership and the Instituto de
Astronom\'{\i}a de la Universidad Nacional Aut\'onoma de M\'exico.
The Apache Point Observatory 3.5 m telescope is owned and operated by
the Astrophysical Research Consortium. We thank the observatory
director, Suzanne Hawley, and site manager, Bruce Gillespie, for their
support of this project.  The Subaru Telescope is operated by the
National Astronomical Observatory of Japan.  The William Herschel
Telescope is operated by the Isaac Newton Group, on the island of La
Palma in the Spanish Observatorio del Roque de los Muchachos of the
Instituto de Astrofisica de Canarias.  Based in part on observations
made with the Nordic Optical Telescope, operated on the island of La
Palma jointly by Denmark, Finland, Iceland, Norway, and Sweden, in the
Spanish Observatorio del Roque de los Muchachos of the Instituto de
Astrofisica de Canarias.  Kitt Peak National Observatory, National
Optical Astronomy Observatory, is operated by the Association of
Universities for Research in Astronomy (AURA), Inc., under cooperative
agreement with the NSF.  The W. M. Keck Observatory is operated as a
scientific partnership among the California Institute of Technology,
the University of California, and NASA; it was made possible by the
generous financial support of the W. M. Keck Foundation.  Based
partially on observations made with the Italian Telescopio Nazionale
Galileo (TNG) operated on the island of La Palma by the Fundaci\'on
Galileo Galilei of the INAF (Istituto Nazionale di Astrofisica) at the
Spanish Observatorio del Roque de los Muchachos of the Instituto de
Astrof\'{\i}sica de Canarias.

This work was supported in part by the Kavli Institute for
Cosmological Physics at the University of Chicago through grants NSF
PHY-0114422 and NSF PHY-0551142, and by an endowment from the Kavli
Foundation and its founder Fred Kavli.  This work was also partially
supported by the US Department of Energy through grants
DE-FG02-08ER41562 to Rutgers University (PI: S.W.J.) and
DE-FG02-08ER41563 to U.C. Berkeley (PI: A.V.F.), as well as by
NSF grants AST-0607485 and AST-0908886 (PI: A.V.F.).


\end{document}